\documentclass[twocolumn]{aastex631}

\usepackage{amsmath,braket,bm,ulem}

\definecolor{darkgreen}{rgb}{0,0.60,0}

\definecolor{customblue}{rgb}{0.3,0.6,1.0}

\begin{document}



\title{The impact of angle-dependent partial frequency redistribution\\ on the scattering polarization of the solar Na~{\sc i} D lines}

\author{Gioele Janett}
\affiliation{Istituto ricerche solari Aldo e Cele Daccò (IRSOL), Faculty of Informatics, Università della Svizzera italiana (USI), CH-6605 Locarno, Switzerland}
\affiliation{Euler Institute, Universit\`a della Svizzera italiana (USI), CH-6900 Lugano, Switzerland}

\author{Ernest Alsina Ballester}
\affiliation{Instituto de Astrof\'isica de Canarias (IAC), E-38205 La Laguna, Tenerife, Spain}
\affiliation{Departamento de Astrof\'{i}sica, Universidad de La Laguna, E-38206 La Laguna, Tenerife, Spain}


\author{Luca Belluzzi}
\affiliation{Istituto ricerche solari Aldo e Cele Daccò (IRSOL), Faculty of Informatics, Università della Svizzera italiana (USI), CH-6605 Locarno, Switzerland}
\affiliation{Leibniz-Institut f\"ur Sonnenphysik (KIS), D-79104 Freiburg i.~Br., Germany}
\affiliation{Euler Institute, Universit\`a della Svizzera italiana (USI), CH-6900 Lugano, Switzerland}

\author{Tanaus\'u del Pino Alem\'an}
\affiliation{Instituto de Astrof\'isica de Canarias (IAC), E-38205 La Laguna, Tenerife, Spain}
\affiliation{Departamento de Astrof\'{i}sica, Universidad de La Laguna, E-38206 La Laguna, Tenerife, Spain}

\author{Javier Trujillo Bueno}
\affiliation{Instituto de Astrof\'isica de Canarias (IAC), E-38205 La Laguna, Tenerife, Spain}
\affiliation{Departamento de Astrof\'{i}sica, Universidad de La Laguna, E-38206 La Laguna, Tenerife, Spain}
\affiliation{Consejo Superior de Investigaciones Científicas, Spain}


\begin{abstract}
The long-standing paradox of the linear polarization signal of the Na~{\sc{i}} D$_1$ line was recently resolved by accounting for the
atom's hyperfine structure and the detailed spectral structure of the incident radiation field.
That modeling relied on the simplifying angle-averaged (AA) approximation for partial frequency redistribution (PRD) in scattering, which potentially neglects important angle-frequency couplings. 
%
This work aims at evaluating the suitability of a PRD--AA modeling for the D$_1$ and D$_2$ lines through comparisons with general angle-dependent (AD) PRD calculations, both in the absence and presence of magnetic fields.
%
We solved the radiative transfer problem
for polarized radiation
in a one-dimensional semi-empirical atmospheric model with microturbulent and isotropic magnetic fields, accounting for PRD effects,
comparing PRD--AA and PRD--AD modelings. 
The D$_1$ and D$_2$ lines are modeled separately  
as two-level atomic system with hyperfine structure.
%
The numerical results confirm that a spectrally structured radiation field induces
linear polarization in the D$_1$ line.
However, the PRD--AA approximation greatly impacts the $Q/I$ shape, producing an antisymmetric pattern instead of the more symmetric PRD--AD one,
while presenting a similar
sensitivity to magnetic fields between $10$ and $200$~G.
Under the PRD--AA approximation, the $Q/I$ profile of the D$_2$ line presents an artificial dip in its core,
which is not found for the PRD--AD case.
%
We conclude that accounting for PRD--AD effects
is essential to suitably model the scattering polarization of the Na~{\sc{i}} D lines.
These results bring us 
closer to exploiting the full diagnostic potential of these lines for the elusive chromospheric magnetic fields.
\end{abstract}

\keywords{Radiative transfer -- Scattering -- Polarization -- Sun: atmosphere -- Methods: numerical}


\section{Introduction}\label{sec:sec1}
Measurements of the polarization of the solar radiation are an essential resource for exploring the properties of the solar atmosphere,
and especially its magnetism. 
High-precision spectro-polarimetric observations performed in quiet regions of the Sun close to the limb reveal numerous and varied linear polarization features
that are collectively referred to as the 
Second Solar Spectrum
\citep[e.g.,][]{stenflo1997,gandorfer2002}. 
These polarization signals are produced by the
scattering of 
anisotropic radiation (i.e., scattering polarization). 
During the scattering process, the radiation 
induces population imbalances and coherence between atomic states that belong to the same level or term (i.e., atomic level polarization). 
The presence of a magnetic field modifies the atomic level polarization, and thus scattering polarization, via the Hanle effect.
Each spectral line is sensitive to the Hanle effect in a given range of field strengths, depending on the Land\'{e} factors and the lifetimes of the levels involved in the spectral line transition \citep[e.g.,][]{trujillo_bueno2001}. 
Nowadays, the power of magnetic field diagnostics based on the joint action of the 
Hanle and Zeeman effects is well established,
especially for obtaining information on photospheric, chromospheric, and coronal 
magnetic fields outside active regions \citep[e.g.,][and references therein]{trujillo_bueno2022}.

A particularly interesting instance of scattering polarization, and the focus of this work, was observed by \citet{stenflo1997}. 
The 
conspicuous linear polarization signal they reported in the core region of Na~{\sc i} D$_1$ at 5896~{\AA} eluded a straightforward explanation because this line, which arises from an atomic transition between an upper and lower level that both have total angular momentum $J = 1/2$, was considered to be intrinsically unpolarizable. 
Subsequent investigations that aimed 
at providing a theoretical explanation for these 
enigmatic signals relied on the fact that sodium has nuclear spin $I = 3/2$ and, hence, 
its fine-structure (FS) levels split into several hyperfine-structure (HFS) levels. 
Accounting for HFS, and assuming that the lower level of D$_1$ (the ground level of sodium) has a
substantial amount of atomic polarization, \cite{landi_degl_innocenti1998} could successfully 
fit the observed D$_1$ polarization signals. 
However, the required 
amount of ground level
atomic polarization was incompatible with the presence of inclined 
magnetic fields stronger than about $0.01$~G, whereas theoretical plasma physics arguments and observations in other spectral lines suggest that far stronger fields should be ubiquitous in the quiet solar chromosphere. 
Finding a resolution to this paradox represented a serious challenge to solar physicists for many years.


Several years later, a breakthrough in the modeling
of the Na~{\sc{i}} D$_1$ line 
was presented by \citet{belluzzi2013}. 
For the unmagnetized case, these authors showed that 
scattering polarization signals of substantial amplitude could be produced without 
the need for ground-level polarization \citep[see also][]{belluzzi2015}. 
It was demonstrated that such linear polarization signals can be explained by taking into account the detailed spectral structure of the incident anisotropic radiation over 
the small wavelength intervals spanned by the HFS transitions that compose the D$_1$ line, so that the various HFS components can possibly be pumped by a slightly different radiation field.
It must be stressed that within the framework of a first-order theory of polarization,
that is considering the limit of complete frequency redistribution (CRD), 
it is not possible to account for this differential pumping together with the impact of coherence between different HFS levels, because this would not comply with the flat-spectrum condition \citep[see][]{landi_deglinnocenti+landolfi2004}.
To accurately model these effects, it is therefore necessary to formulate the problem using high-order approaches, which include partial frequency redistribution (PRD) effects (i.e., frequency correlations between incident and scattered radiation).
The suitability of this explanation was confirmed when
\citet{alsina_ballester2021}
modeled the polarization of the solar sodium 
doublet radiation 
accounting for both collisions and magnetic fields, 
within the framework of the quantum theory of spectral line polarization described in~\citet{bommier2017}. 
They also highlighted the sensitivity of these polarization signals to the presence of magnetic fields in the gauss range, opening up a new window for probing the elusive magnetic fields of the 
solar chromosphere in the present new era of large-aperture solar telescopes. 

Because of the high computational requirements of modeling these scattering polarization signals, \citet{alsina_ballester2021}
made the so-called angle-averaged (AA) simplifying assumption 
in their calculations
(see, e.g., \citealt{rees1982}; \citealt{sampoorna2014} and references therein; 
\citealt{belluzzi2014}; \citealt{alsinaballester2017}), 
which accounted for the HFS of sodium and for the quantum interference between states belonging to different FS 
levels of the upper term (i.e., $J$-state interference) and between states belonging to different HFS levels of the same HFS level (i.e., $F$-state interference).
As summarized in Section~\ref{sec:AA_approximation}, the AA approximation smears out geometrical dependencies of the problem, and can, in principle, introduce significant inaccuracies in the synthesized profiles \citep[see, e.g.,][]{janett2021a}.

The aim of this work is to evaluate the suitability of angle-averaging in this context through a comparison with angle-dependent (AD) PRD calculations, which are more accurate but have a significantly higher numerical cost. 
For this goal, we carried out a series of calculations
with a code capable of solving the 
non-local-thermodynamical-equilibrium (non-LTE)
RT problem in static 1D semi-empirical models of the solar atmosphere for 
two-level model atoms (in which $J$-state interference are neglected)
with HFS, taking scattering polarization and PRD--AD effects into 
account. 

The article is organized as follows: Section~\ref{sec:transfer}
exposes the considered RT problem for polarized radiation 
and outlines the adopted solution strategy.
Section~\ref{sec:atomic_atmos_models} describes the considered
atomic and atmospheric models.
In Section~\ref{sec:numerical_results}, we report and analyze
the synthetic emergent Stokes profiles of the Na~{\sc{i}} D$_1$ and D$_2$ lines,
comparing PRD--AA and PRD--AD calculations.
Finally, Section~\ref{sec:conclusions} provides remarks and conclusions. 

\section{Transfer problem for polarized radiation}
\label{sec:transfer}
The intensity and polarization of a radiation beam are fully described by the four Stokes parameters $I$, $Q$, $U$, and $V$.
The Stokes parameter $I$ quantifies the intensity, $Q$ and $U$ jointly quantify the linear polarization, while $V$ quantifies the circular polarization \citep[e.g.,][]{landi_deglinnocenti+landolfi2004}. 
Hereafter, we will assume stationary conditions so that all the considered quantities are time-independent.

The intensity and polarization of a radiation beam propagating in a medium (e.g., the plasma of a stellar atmosphere) 
change as the radiation interacts with the particles therein.
This modification is fully described by the RT equation for polarized radiation,
which is a system of coupled first-order, inhomogeneous, ordinary differential equations.
Defining the Stokes parameters as the four components of the Stokes vector $\bm{I}=(I_1,I_2,I_3,I_4)^T=(I,Q,U,V)^T\in \mathbb{R}^4$, the RT equation
can be written as
\begin{equation}
 	\vec{\nabla}_{{\bm{\Omega}}}\bm{I}(\bm{r},\bm{\Omega},\nu) = 
	- K(\bm{r},\bm{\Omega},\nu) \, \bm{I}(\bm{r},\bm{\Omega},\nu) + 
	\bm{\varepsilon}(\bm{r},\bm{\Omega},\nu),
	\label{eq:rte}
\end{equation}
where $\bm{r}$ denotes the spatial point,
$\nu$ the radiation frequency, 
%
and $\vec{\nabla}_{\bm{\Omega}}$ denotes the spatial derivative
along the direction specified by the unit vector $\bm{\Omega}=(\theta,\chi)$,
$\theta$ and $\chi$ being the polar angles (inclination and azimuth, respectively)\footnote{We consider a right-handed Cartesian reference system with the $z$-axis directed along the local vertical. The inclination $\theta$ is the angle between the $z$-axis and $\bm{\Omega}$ (this angle corresponds to the heliocentric angle of the observed point), while the azimuth $\chi$ is the angle (measured counter-clockwise) between the $x$-axis and the projection of $\bm{\Omega}$ on the $xy$-plane.}. 
The propagation matrix $K \in \mathbb{R}^{4 \times 4}$ is given by
\begin{equation}
	K = \begin{pmatrix}
		\eta_1 & \eta_2 & \eta_3 & \eta_4 \\
		\eta_2 & \eta_1 & \rho_4 & -\rho_3 \\
		\eta_3 & -\rho_4 & \eta_1 & \rho_2 \\
		\eta_4 & \rho_3 & -\rho_2 & \eta_1 \\
	\end{pmatrix}. 
\label{eq:prop_matrix}
\end{equation}
The elements of $K$ describe absorption ($\eta_1$), dichroism ($\eta_2$, $\eta_3$, and $\eta_4$), and anomalous dispersion ($\rho_2$, $\rho_3$, and $\rho_4$) phenomena.
The emission vector $\bm{\varepsilon}=(\varepsilon_1,\varepsilon_2,\varepsilon_3,\varepsilon_4)^T\in \mathbb{R}^4$ describes the radiation emitted by the plasma in the four Stokes parameters. 

In the frequency interval of a given spectral line, the elements of $K$ and $\bm{\varepsilon}$ depend on the state of the atom (or molecule) giving rise to that line.
In general, this state has to be determined by solving a set of rate equations (statistical equilibrium equations), which describe the interaction of the atom with the 
radiation field (radiative processes), other particles present in the plasma (collisional processes), and external magnetic and electric fields.
The emission vector, in particular, can be written as the sum of two terms, namely 
\begin{equation}\label{eq:epsilon_tot}
	\bm{\varepsilon}(\bm{r},\bm{\Omega},\nu) =
	\bm{\varepsilon}^{\mathrm{sc}}(\bm{r},\bm{\Omega},\nu) +
	\bm{\varepsilon}^{\mathrm{th}}(\bm{r},\bm{\Omega},\nu), 
\end{equation}
where $\bm{\varepsilon}^{\mathrm{sc}}$ describes the contribution from atoms that are radiatively excited (scattering term), and $\bm{\varepsilon}^{\mathrm{th}}$ describes the contribution due to atoms that are collisionally excited (thermal term).
We note that, in the solar atmosphere,
stimulated emission is negligible
in the spectral interval around the Na~{\sc i} D lines
and it is consequently neglected.
For certain, relatively simple, atomic models (for instance a two-term atomic system with an unpolarized lower term whose levels are infinitely sharp), an analytic solution exists for the statistical equilibrium equations.  
In such scenarios, the scattering contribution to the emission vector can be directly related to the radiation field that illuminates the atom (incident radiation) through the redistribution matrix formalism. 
Following the convention that primed and unprimed quantities refer to the incident and scattered radiation, respectively, it can be written in terms of the so-called scattering integral
\begin{equation}\small
	\bm{\varepsilon}^{\mathrm{sc}}(\bm{r},\bm{\Omega},\nu) = k_L(\bm{r})
	\!\!\int \!\! \mathrm{d} \nu^\prime\!\!  \oint 
	\frac{\mathrm{d} \bm{\Omega}^\prime}{4 \pi} R(\bm{r},\bm{\Omega}^\prime,\bm{\Omega},\nu^\prime,\nu ) \, \bm{I}(\bm{r},\bm{\Omega}^\prime,\nu^\prime),
	\label{eq:scat_int}
\end{equation}
where the factor $k_L$ is the frequency-integrated absorption coefficient, and $R \in \mathbb{R}^{4 \times 4}$ is the redistribution matrix.
The element $R_{ij}(\bm{r},\bm{\Omega}^\prime,\bm{\Omega},\nu^\prime,\nu )$ of the redistribution matrix relates the $i$-th Stokes component of the emissivity, in direction $\bm{\Omega}$ and at frequency $\nu$, to the $j$-th Stokes component of the incident radiation with direction $\bm{\Omega}^\prime$ and frequency $\nu^\prime$.

For an atomic system with infinitely sharp lower states, as the one considered in this work (see Section 3 for more details), the redistribution matrix can be separated into
{\small
\begin{equation*}
	R(\bm{r},\bm{\Omega}^\prime,\bm{\Omega},\nu^\prime,\nu) = 
	R^{\scriptscriptstyle \mathrm{II}}(\bm{r},\bm{\Omega}^\prime, \bm{\Omega},\nu^\prime, \nu) + 
	R^{\scriptscriptstyle \mathrm{III}}(\bm{r},\bm{\Omega}^\prime, \bm{\Omega}, \nu^\prime, \nu),
\end{equation*}}
\noindent where the notation introduced in \cite{hummer1962} has been used. The $R^{\scriptscriptstyle \mathrm{II}}$ matrix quantifies scattering processes that are coherent in the rest frame of the atom and $R^{\scriptscriptstyle \mathrm{III}}$ quantifies those that are totally incoherent due to the impact of elastic collisions.
The expressions for $R^{\scriptscriptstyle \mathrm{II}}$ and $R^{\scriptscriptstyle \mathrm{III}}$ in the observer's frame (i.e., including the Doppler shifts due to thermal atomic motions) in the case of a two-term atom can be obtained from Equation~(A.1) of \cite{bommier2018}.
The same redistribution matrices, particularized to the case of a Maxwellian distribution of atomic velocities (without a bulk component)
under the AA approximation (see Section~\ref{sec:AA_approximation}),
as well as the $K$ matrix elements,
can be found in \cite{alsina_ballester2022}.
The thermal contribution to the emissivity $\bm{\varepsilon}^{\mathrm{th}}$ can be found, for instance, in \cite{alsina_ballester2022b}. 
In this work, we also include the contribution brought by continuum processes. In the considered spectral region, they only contribute to the diagonal elements of $K$. The continuum emissivity is calculated including both the thermal contribution (assumed to be isotropic and unpolarized) and the scattering one (under the assumption of coherent scattering in the observer's frame). More details on the continuum terms can be found in Sect.~2.5 of \citet{alsina_ballester2022}.

Solving the whole RT problem consists in finding a self-consistent solution for the RT equation (\ref{eq:rte}) and the equation for the scattering contribution to the emissivity (\ref{eq:scat_int}).
This problem is in general non-linear because of the factor $k_L$ 
that appears both in the elements of the propagation matrix and in the emission coefficients. 
This factor is proportional to the population of the lower level, which in turn depends non-linearly on the incident radiation field through the statistical equilibrium equations.

We linearize the problem with respect to $\bm{I}$, by fixing \textit{a priori} 
the population of the lower level, and thus the factor $k_L$. In this scenario,
whose suitability is discussed in \cite{janett2021b} and \cite{benedusi2021}, the propagation matrix $K$ and the thermal contribution to the emissivity $\bm{ \varepsilon}^{\mathrm{th}}$ are independent of $\bm{I}$, while the scattering term $\bm{\varepsilon}^{\mathrm{sc}}$ depends on it linearly through the scattering integral shown in Equation~\eqref{eq:scat_int}. 
The population of the lower level can be taken either from the atmospheric model (if provided) or from independent calculations. 
The latter can be carried out with available non-LTE RT codes that 
neglect polarization (which is expected to have a minor impact on the population of ground or metastable levels), but allow considering multi-level atomic models. 
In this way, accurate estimates of the lower level population can be used
\citep[e.g.,][]{janett2021a,alsina_ballester2021}.

\subsection{Angle-averaged approximation}
\label{sec:AA_approximation}

Using the formalism of the irreducible spherical tensors for polarimetry
\citep[e.g., Chapt.~5 of][]{landi_deglinnocenti+landolfi2004},
the ${R}^{{\scriptscriptstyle \rm II}}$ and ${R}^{{\scriptscriptstyle \rm III}}$ redistribution matrices can be decomposed as
{\small
\begin{equation*}
    R^{\scriptscriptstyle X}\!(\bm{r},\bm{\Omega}^\prime\!,\bm{\Omega},\nu^\prime\!,\nu)\! = \!\!\!
    \sum_{K K^\prime Q}\!\!\!
    \mathcal{R}^{{\scriptscriptstyle X},K K^\prime}_{Q}\!(\bm{r},\bm{\Omega}^\prime,\bm{\Omega},\nu^\prime\!,\nu)
    P^{K K^\prime}_{Q}\!(\bm{r},\bm{\Omega}^\prime,\bm{\Omega}),
\end{equation*}}
\noindent with $X = \{\mathrm{II}, \mathrm{III}\}$.
The scattering phase matrix $P$ is independent of frequency and its expression can be found, e.g., in \citet[][]{alsina_ballester2022}.  
The $\mathcal{R}^{{\scriptscriptstyle X}}$ redistribution function locally couples all frequencies and directions of the incident and scattered radiation, making the problem computationally very demanding.
In the absence of bulk velocities, the angular dependence of this function is fully contained in the scattering angle 
\begin{equation*}
\Theta=\arccos\left(\bm{\Omega}^\prime \cdot \bm{\Omega}\right) . 
\end{equation*}
Approximated forms of the functions $\mathcal{R}^{{\scriptscriptstyle X}}$ are frequently applied to simplify the calculations. 
A common one for $\mathcal{R}^{{\scriptscriptstyle \rm II}}$ is the so-called AA, which consists in averaging this function over the scattering angle $\Theta$, namely
{\small
\begin{equation}\label{eq:angle_averaged}
	\mathcal{R}^{{\scriptscriptstyle \rm II-AA},KK^\prime}_{Q}\!(\bm{r},\nu^\prime, \nu) = \frac{1}{2}\!\int_0^\pi \!\!\! {\rm d} {\Theta}\sin(\Theta) \, 
	\mathcal{R}^{{\scriptscriptstyle \rm II},KK^\prime}_{Q}\!(\bm{r},\bm{\Omega}^\prime,\bm{\Omega},\nu^\prime,\nu).
\end{equation}}
\noindent When the AA redistribution function~\eqref{eq:angle_averaged} is used, frequencies and directions are completely decoupled in $\mathcal{R}^{{\scriptscriptstyle \rm II}}$, allowing for a drastic reduction of the computational cost.

\citet{faurobert1987,faurobert1988} and \citet{sampoorna2011,sampoorna2017}
analyzed the suitability of 
the AA approximation in the modeling of polarization signals
in academic scenarios,
concluding that this approximation
introduces relevant inaccuracies in the modeling of the 
linear polarization signals in the core of spectral lines.
More recently, \citet{janett2021a} showed that the AA
approximation introduces an artificial trough in the line-core peak of the $Q/I$ and $U/I$ scattering
polarization profiles of the Ca~{\sc i} 4227 line. 
This highlights the need to evaluate the impact of this 
approximation when modeling the scattering polarization of
lines where PRD effects play a crucial role, like for the D
lines of neutral sodium. 

\citet{leenaarts2012} proposed a generalization of the AA approximation suited for dynamic scenarios, which proved to be highly effective for the unpolarized case. 
This version of the AA approximation sacrifices information on the anisotropy of the radiation field, and
it is thus not suited for the description of scattering polarization. 
However, the AA approximation for the polarized case can be also used in dynamic scenarios by applying the comoving 
frame method for treating bulk velocities \citep[see][and references therein]{sampoorna2016}.
We note that the radiative transfer equation in the comoving frame includes an additional term containing
a derivative w.r.t. frequency, yielding a partial differential equation, which is significantly
more complicated to solve \citep[see, e.g.,][p. 492]{mihalas1978}. 

As far as 
$\mathcal{R}^{{\scriptscriptstyle \rm III}}$ is concerned, a widely used approximation is to assume that the scattering processes described by this matrix are totally incoherent also in the observer's frame
\citep[e.g.,][]{mihalas1978,bommier1997b,alsinaballester2017,benedusi2022}.
This assumption is also applied in this work.

\subsection{Numerical solution strategy}
\label{sec:solution_strategy}

Following the works by \citet{janett2021b} and \citet{benedusi2021,benedusi2022}, we first present an algebraic formulation of the considered linearized RT problem for polarized radiation.
Starting from this formulation, we then apply a parallel solution strategy, based on Krylov iterative methods with physics-based preconditioning.
This strategy allows us to routinely solve the problem in semi-empirical 1D models of the solar atmosphere, considering the angle-dependent  expression of $R^{\scriptscriptstyle \rm II}$.

The problem is discretized by introducing suitable grids
for the continuous variables $\bm{r}$, $\theta$, $\chi$, and $\nu$
(see Section~\ref{sec:numerical_setting}).
Provided the radiation field at all nodes on the angular grid $(\theta_l,\chi_m)$ ($l=1,...,N_\theta$, $m=1,...,N_\chi$) and frequency grid $\nu_j$ ($j=1,...,N_\nu$), for a given spatial grid point $\bm{r}_i$ ($i=1,....,N_r$), the scattering integral~\eqref{eq:scat_int} is evaluated in terms of suitable angular and spectral quadratures.
Provided the propagation matrix \eqref{eq:prop_matrix}
and the emission coefficients~\eqref{eq:epsilon_tot} at all spatial points
$\bm{r}_i$, for a given direction $(\theta_l,\chi_m)$ and frequency $\nu_j$, the RT equation \eqref{eq:rte} is numerically solved
by applying the $L$-stable DELO-linear method combined with a linear conversion to optical depth \citep[see, e.g.,][]{janett2017a,janett2018a,janett2018b}.
In Equation~\eqref{eq:rte} we impose the following boundary conditions: 
we assume that no radiation is entering from the upper boundary, while the radiation entering the atmosphere from the lower boundary is assumed isotropic, unpolarized, and equal to the Planck function.

After introducing the collocation vectors
$\mathbf{I}\in\mathbb R^N$ and $\bm{\epsilon}\in\mathbb R^N$ $(N=4 N_r \, N_\nu \, N_\theta \, N_\chi)$,
which contain the numerical approximations
of the Stokes vector and the emission vector, respectively, at all $N$ nodes,
the whole discrete RT problem can be then formulated in the compact form
given by
\citep[see][]{janett2021a,benedusi2022}
\begin{equation}
	(Id - \Lambda \Sigma) \mathbf{I} = \Lambda \bm{\epsilon}^{\mathrm{th}} + \mathbf{t} .
	\label{eq:lin_sys}
\end{equation}
In this linear system, 
$Id\in\mathbb R^{N\times N}$ is the identity matrix.  
$\Lambda: \mathbb{R}^N \rightarrow \mathbb{R}^N$ is the transfer operator, which encodes the formal solver and the propagation matrix.
The scattering operator, which
encodes the numerical evaluation of the scattering integral~\eqref{eq:scat_int},
is given by the sum of different components, namely,
\begin{equation*}
	\Sigma = \Sigma^{\scriptscriptstyle \rm II} + 
	\Sigma^{\scriptscriptstyle \rm III} + 
	\Sigma^{\rm c} ,
\end{equation*}
where $\Sigma^{\scriptscriptstyle \rm II}$ and $\Sigma^{\scriptscriptstyle \rm III}$ encode the contributions from $R^{\scriptscriptstyle \rm II}$ and $R^{\scriptscriptstyle \rm III}$, respectively, and $\Sigma^{\rm c}$ the
scattering contribution from the continuum.
The vector $\bm{\epsilon}^{\rm th} \in \mathbb{R}^N$ encodes the thermal emissivity, and the vector $\mathbf{t}\in\mathbb R^N$ encodes the boundary conditions.

\begin{figure}[ht!]
    \centering
    \includegraphics[width=0.49\textwidth]{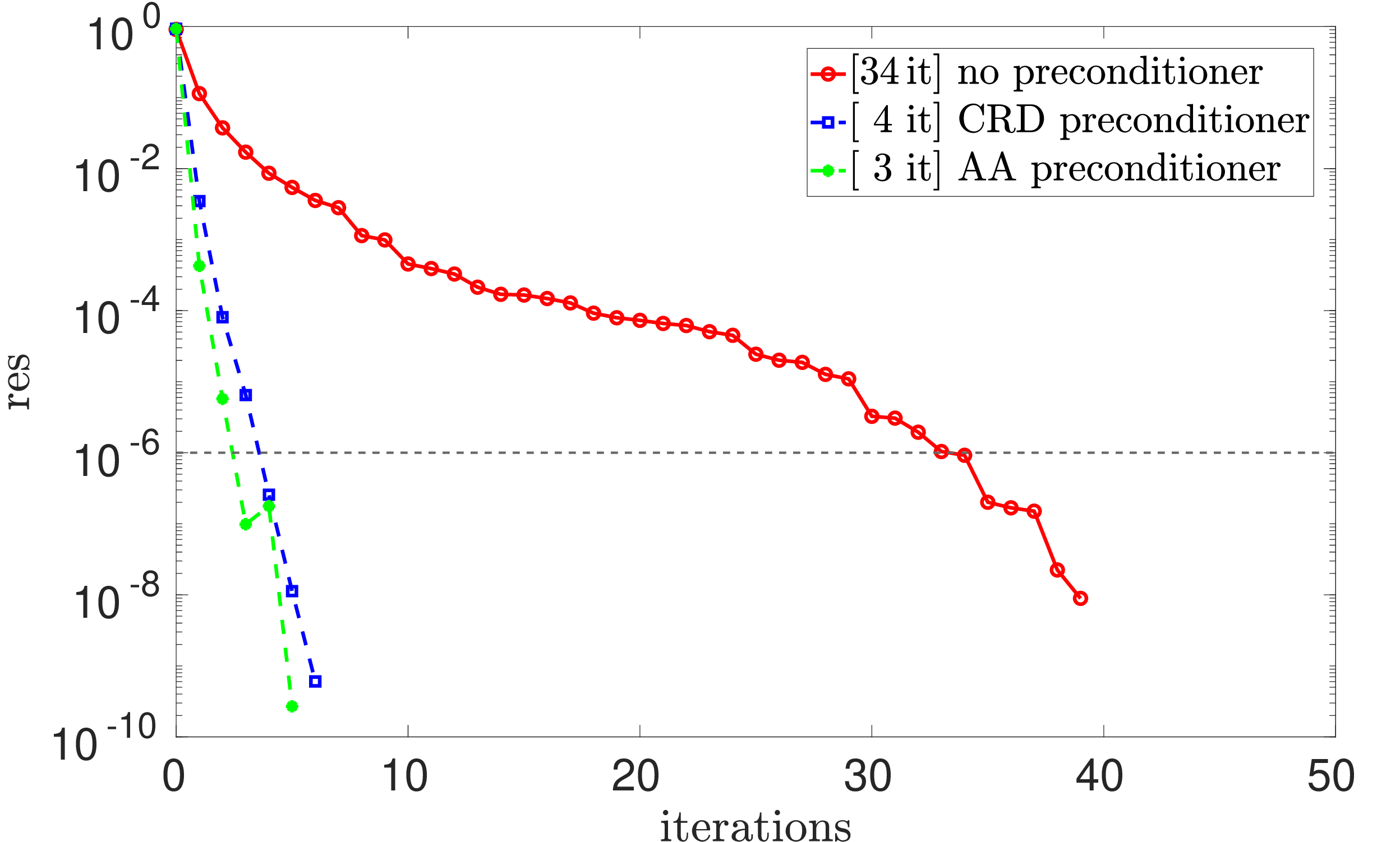}
    \caption{Convergence history for the GMRES iterative solution of~\eqref{eq:lin_sys} for the Na~{\sc{i}} D$_1$ line, exploiting none, CRD, and PRD--AA preconditioners.
    The number of iterations to convergence are reported in square brackets.
    The horizontal dashed line represents the tolerance (${\rm tol}=10^{-6}$) for the preconditioned relative residual (${\rm res}$).}
    \label{fig:convergence_naiD1}
\end{figure}

Under the assumption that $k_L$ is known 
\textit{a priori} \citep[see, e.g.,][]{janett2021a}, the operators $\Lambda$ and $\Sigma$ do not depend on $\mathbf{I}$, and the problem~\eqref{eq:lin_sys} is thus linear.
The right-hand side vector $\Lambda \bm{\epsilon}^{\mathrm{th}} + \mathbf{t}$
is computed \textit{a priori}, while
the action of the matrices $\Lambda$ and $\Sigma$ is encoded in a matrix-free form.
We solve the linear system \eqref{eq:lin_sys} by applying a matrix-free, preconditioned GMRES method.
Figure~\ref{fig:convergence_naiD1} illustrates the convergence history for preconditioned and unpreconditioned GMRES iterative methods
applied to~\eqref{eq:lin_sys},
by modeling the Na~{\sc{i}} D$_1$ line via the numerical setting described
in Section~~\ref{sec:numerical_setting}, in the absence of magnetic fields.
In particular, we exploit two different preconditioners based on two simplified descriptions of the scattering process:
(i) by considering the limit of CRD and (ii) by applying the AA approximation.
Both CRD- and PRD--AA-based preconditioners have a noticeable impact. 
We note that the computational cost for the former is substantially lower than for the latter.
In the calculations presented in this paper, preconditioning is thus performed by describing scattering processes in the limit of CRD.
The preconditioned GMRES solvers outperform the unpreconditioned one, strongly reducing the number of iterations to convergence. 
The reader is referred to \citet{benedusi2022} for more details on this solution strategy.

\section{Atomic and atmospheric models}\label{sec:atomic_atmos_models}

In this section, we present the atomic and the atmospheric models used in the calculations.
%
%
A strictly correct RT modeling of the scattering polarization of the 
Na~{\sc i} D lines
should be carried out considering 
a two-term atom with HFS \citep[see][]{alsina_ballester2021}.
The upper term ($^2\mathrm{P}^o$)
consists of two FS $J$ levels,
having total angular momentum $J = 1/2$ (upper level of D$_1$)
and $J = 3/2$ (upper level of D$_2$).
The lower term ($^2\mathrm{S}$) consists of the ground level of sodium 
with $J = 1/2$, which is the lower level of both D$_1$ and D$_2$.
Because this level has a very long lifetime, 
all its sublevels can be taken to be infinitely sharp.
The lower level can also be treated as unpolarized; because of its long lifetime, it is strongly depolarized by magnetic fields of just a few tens 
of milligauss \citep{landi_degl_innocenti1998} and by elastic collisions with neutral hydrogen,
for densities as low as $n_{\mathrm{H}} \sim 10^{14}$cm$^{3}$ \citep[see][]{kerkeni2002}. 
Due to the interaction with the nuclear spin ($I = 3/2$), the $J$ levels split into various HFS $F$ levels. 

In the presence of external magnetic fields such that
the splitting between magnetic sublevels is comparable to the separation
between HFS or FS levels,
their energies must be calculated in the incomplete Paschen-Back (IPB) effect regime,\footnote{It is common in the literature 
to refer to the Paschen-Back effect for HFS as the Back-Goudsmit effect.}
accounting for the related mixing between eigenstates of total angular momentum.

Because of the large energy separation between the upper $J$ levels of D$_1$ and D$_2$, the $J$-state interference between them is not appreciably modified by the presence of a magnetic field, and it only affects the scattering polarization pattern outside the line core regions.
This can be seen from the similarities between the synthetic $Q/I$ profiles shown in Figure 4 of \citet{belluzzi2013} and in Figure 1 of \citet{belluzzi2015}, in which $J$-state interference was taken into account and neglected, respectively. We also verified this using the more recent numerical code described in
\citet{alsina_ballester2022}, which is suitable for modeling lines that arise from a two-term atomic system with HFS, in the presence of arbitrary magnetic fields, under the AA approximation.

\subsection{Two-level atom with HFS for Na~{\sc i} D lines}



At the time of writing, no numerical code exists that
can handle
the computational complexity of
the non-LTE RT problem for polarized radiation for a two-term atom with HFS, while accounting for PRD effects in the general AD case.
However, bearing in mind that $J$-state interference does not significantly modify the line-core polarization signals of the Na~{\sc i} D lines, for the purposes of this work we find it suitable to model the D1 and D2 lines separately, considering a two-level atom with HFS for each of them.
A similar strategy was followed by
\citet{sampoorna2019} and \citet{nagendra2020}
in their PRD-AA vs PRD-AD investigation of the 
Na {\sc i} D$_2$ line in isothermal slab models.
The D$_1$ (D$_2$) line is modeled setting $J_\ell=1/2$ and $J_u=1/2 \,\,(3/2)$, and taking into account the HFS due to a nuclear spin $I=3/2$.
The calculations are carried out using existing codes for a two-term atom without HFS, and applying them to the formally equivalent case of a two-level atom with HFS 
\citep[e.g.,][]{landi_deglinnocenti+landolfi2004}.
In the unmagnetized case, the equivalence between a two-term atom and a two-level atom with HFS is obtained with the following formal substitutions in the quantum numbers: 
$S \rightarrow I$, $L \rightarrow J$, $J \rightarrow F$ (see also Eq.~\eqref{eq:FS_to_HFS}),
where $S$ and $L$ have the usual meaning of total electronic spin and orbital angular momentum quantum numbers, respectively.
The generalization to the magnetic case is described in Appendix~\ref{app:twolevel_HFS}.
We used experimental data for the energies of the upper $J$ levels \citep{kramida2022}.
The energy splittings of the HFS $F$ levels were calculated using recent experimental values for the HFS constants 
for the various $J$ levels \citep{steck03}.\footnote{
 That work used the American convention for the HFS constants, in contrast to the expressions used in Appendix~\ref{app:twolevel_HFS}. A suitable conversion was thus applied 
 for this work. 
}

\subsection{1D semi-empirical atmospheric model}
\label{sec:atmospheric_model}

In this work, we consider the 1D semi-empirical solar atmospheric model~C of \citet{fontenla1993} 
either in the absence or in the presence of microturbulent magnetic fields (see Section~\ref{sec:microB}), taking the micro-turbulent
velocities as determined semi-empirically in \cite{fontenla1991}. 
In a 1D model, the spatial dependency of the physical quantities entering the 
RT problem is fully described by the height coordinate $z$, which replaces the coordinate $\bm{r}$.
In this work we do not consider deterministic magnetic fields and bulk velocities. The problem is thus characterized by axial symmetry,
which allow us to 
keep computational costs
down to a manageable level. 
It can be expected that the impact of the general AD treatment of PRD effects will be greater when the radiation field scattered by the atom has a more complex angular dependence, as happens when the complex 3D geometrical structure of the solar chromospheric plasma is taken into account
\citep[e.g.,][and references therein]{benedusi2023,anusha2023}.
\subsection{Microstructured magnetic fields}
\label{sec:microB}

In the solar atmosphere, 
it is common to find magnetic fields that vary
on spatial scales below the resolution element of standard observations, which are often referred to as tangled \citep[e.g.][]{trujillo_bueno2004,manso_sainz2004}.
Throughout this work, we consider \textit{unimodal microstructured 
isotropic magnetic fields}
\citep[see][]{stenflo1994,trujillo_bueno1999,alsinaballester2017},
that is, fields with a fixed strength but whose orientation
changes 
at scales below the mean free path of photons, 
such that they are uniformly distributed over all directions.
The expressions for the RT 
coefficients (i.e., elements of the propagation matrix and components of the emission vector) 
in the presence of such tangled magnetic fields are
discussed in \citet{alsina_ballester2022}.

\newpage

\section{Numerical results}\label{sec:numerical_results}

In this section, we provide quantitative results
on the suitability of a PRD--AA modeling for the Na~{\sc{i}} D lines,
through comparisons with the general PRD--AD calculations,
both in the absence and presence of magnetic fields.
As discussed 
in Section~\ref{sec:atomic_atmos_models},
the D$_1$ and D$_2$ lines are modeled separately,
considering for each of them a
two-level system with HFS,
in which the Paschen-Back effect is taken into account for the magnetic splitting. 
This section focuses on the modeling of the D$_1$ line at 5896~{\AA},
whereas the modeling of the D$_2$ line at 5890~{\AA} is reported in
Appendix~\ref{app:D2}.

\subsection{Numerical setting}
\label{sec:numerical_setting}

The wavelength interval
$[\lambda_{\min},\lambda_{\max}]=[5894.4\text{\,\AA},5897.4\text{\,\AA}]$
is discretized with ${N_\nu=161}$ logarithmically distributed nodes. 
This frequency grid, chosen to suitably sample the spectral line under investigation,
is denser in the line core and coarser in the wings.
The spatial grid is provided by the considered
1D semi-empirical atmospheric model C of \citet{fontenla1993},
which discretizes the height interval $[z_{\min},z_{\max}]=[-100~\text{km},2219~\text{km}]$ with $N_r=70$
unevenly distributed spatial nodes.
We adopted a very common tensor product angular grid
with $N_\chi=12$ and $N_{\theta}=12$,
for a total of $N_\Omega=144$ nodes,
which guarantees  
the accurate calculation of the scattering polarization signals
(see Appendix~\ref{app:angular_grid} for details). 

The proposed iterative solver consists
of two nested GMRES iterations \citep[see][]{benedusi2022}.
We use no restart and the same threshold tolerance
${\rm tol}$ for both GMRES iterations.
As a stopping criterion for the iterative method,
we adopt the inequality
$${\rm res}<{\rm tol}=10^{-6},$$
in which the preconditioned relative residual ${\rm res}$ 
is a scalar
that indicates the accuracy of the approximate solution. 
We adopt a zero initial guess.


\begin{figure}[ht!]
    \centering
    \includegraphics[width=0.49\textwidth]{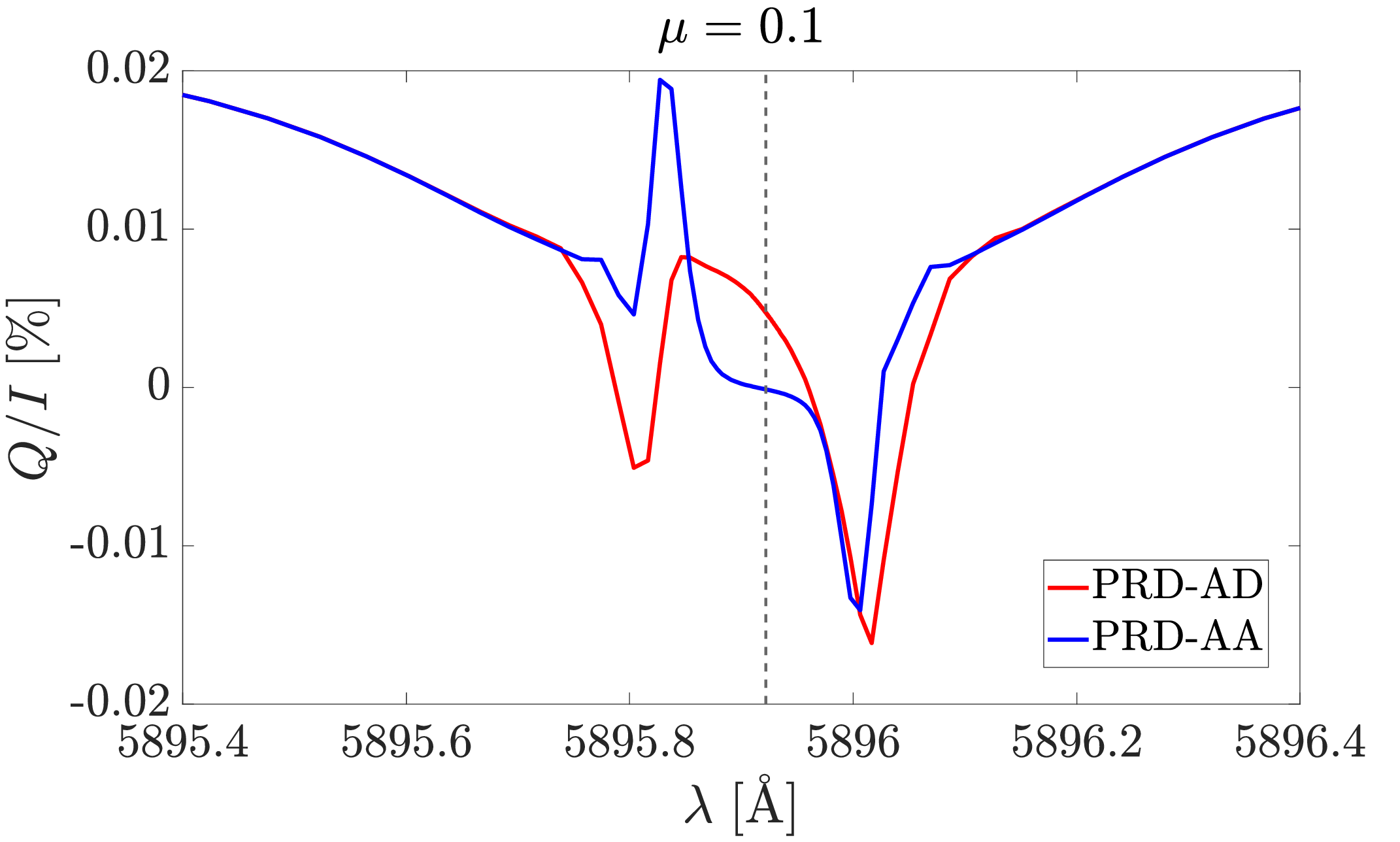}
    \caption{
    Emergent $Q/I$ profiles for the Na~{\sc{i}} D$_1$ line at $\mu = 0.1$, calculated with a two-level model atom and in the atmospheric model C of \cite{fontenla1993}, without a magnetic field. 
    The results of calculations are shown
    taking PRD effects into account both in the general AD case (red curve)
    and under the AA approximation (blue curve),
    in both cases neglecting the impact 
    of $J$-state interference.
    The vertical dashed line indicates the line-center wavelength.
    }
    \label{fig:B0}
\end{figure}

\subsection{Impact of PRD--AD treatment on D$_1$ polarization}

The intensity profiles of PRD--AA and PRD--AD calculations are essentially identical in all the considered cases, and thus they are not explicitly shown.
Figure~\ref{fig:B0} shows the $Q/I$ patterns of the D$_1$ line obtained for PRD--AA and 
for PRD--AD calculations
in the absence of magnetic fields,
using the numerical scheme described in the previous sections. 
In order to have scattering polarization signals with a relatively large amplitude in the considered geometry,
we show a significantly inclined line of sight with $\mu = \cos(\theta) = 0.1$.
The PRD--AA calculation yields a largely antisymmetric $Q/I$ profile around the D$_1$ line core, 
with a positive peak on the blue side of the line center, and a negative one on the red side.
We find a {very good} qualitative
agreement between this pattern and 
the one reported in Figure~4 of \cite{belluzzi2013}, 
although 
a strict comparison cannot be made because different formal solvers were used.
Likewise, it also presents a good qualitative agreement with the polarization signals in the D$_1$ core region shown in Figure~1 of \cite{belluzzi2015} and in Figure~2 of \cite{alsina_ballester2021}.
As expected, the $Q/I$ lobes outside the line core
reported in the latter two investigations
cannot be reproduced with the two-level atomic model considered in
the present work,
because such lobes arise from $J$-state interference. 

The PRD--AD calculation yields a scattering polarization
profile that presents
substantial differences in the shape
with respect to the PRD--AA calculation. 
Whereas a similar red negative peak is obtained
for both treatments, 
the positive blue peak 
that was found in the PRD--AA case is shifted toward the line center for the PRD--AD calculation and a negative peak is found in its place. 
Thus, instead of the antisymmetric 
pattern resulting from the PRD--AA calculation, the PRD--AD calculation produces a far more symmetric profile, with a positive $Q/I$ peak near the line center
and negative peaks at $\sim0.1$~\AA~to the red and blue of it. 
We also note that the positive $Q/I$ peak found in the PRD--AD case does not fall exactly at line center, but  instead its maximum is slightly shifted to the blue.
When artificially neglecting the HFS splitting in the lower level, the resulting $Q/I$ profile only shows a depolarization feature in the line core.
This confirms that the physical origin of the line scattering polarization is the same as for PRD--AA calculations,
that is the variation of the anisotropy of the radiation field over the spectral range spanned by the HFS transitions of the D$_1$ line, as shown by \citet{belluzzi2013}.
Even though the linear polarization profile resulting from AD calculations is rather symmetric, in contrast to the mostly antisymmetric $Q/I$
profile obtained from AA calculations, it is noteworthy that the signal obtained from integrating the former over frequency
is not greater than the one obtained when integrating the latter (see Appendix~\ref{app:QI_integrated}).

Thus, a suitable modeling of the observed linear polarization patterns of the Na~{\sc{i}} D$_1$ line requires making a PRD--AD treatment of scattering 
processes, while considering a two-term atomic model (in order to account for the $J$-state interference) with HFS. 
We recall that an antisymmetric $Q/I$ profile
of the Na~{\sc{i}} D$_1$, obtained from an observation with low temporal and spatial resolution, could be fit remarkably well considering PRD-AA calculations
\citep[see][]{alsina_ballester2021}.
We may expect that such antisymmetric profiles can be modeled also considering AD-PRD effects, once we additionally account for
relevant phenomena such as the spectral smearing of typical instruments and local effects related to the 
three-dimensional and dynamic nature of the solar atmosphere.
An investigation of these effects is, however, beyond the scope of
both the aforementioned investigation and the present one.
%
We also point out that new spectro-polarimetric observations have been recently acquired in the spectral range around the Na~{\sc{i}} D$_1$ line \citep[e.g.,][]{bianda2019}. Such observations present a wide variety of profiles,
a few of which are quite symmetric,
with a shape in
good qualitative agreement with the theoretical PRD--AD profiles presented above.

\begin{figure}[!ht]
    \centering
    \includegraphics[width=0.49\textwidth]{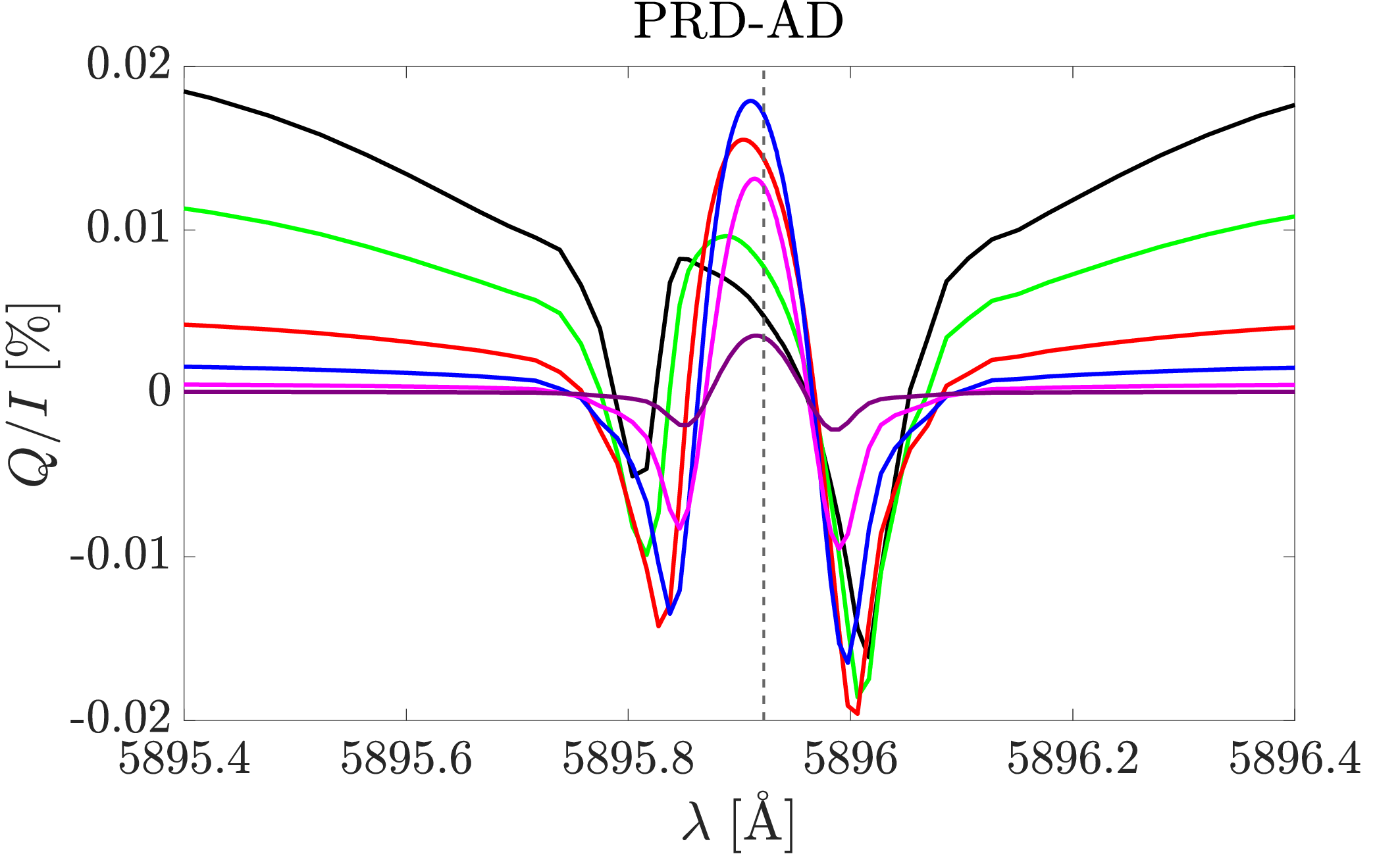}
    \includegraphics[width=0.49\textwidth]{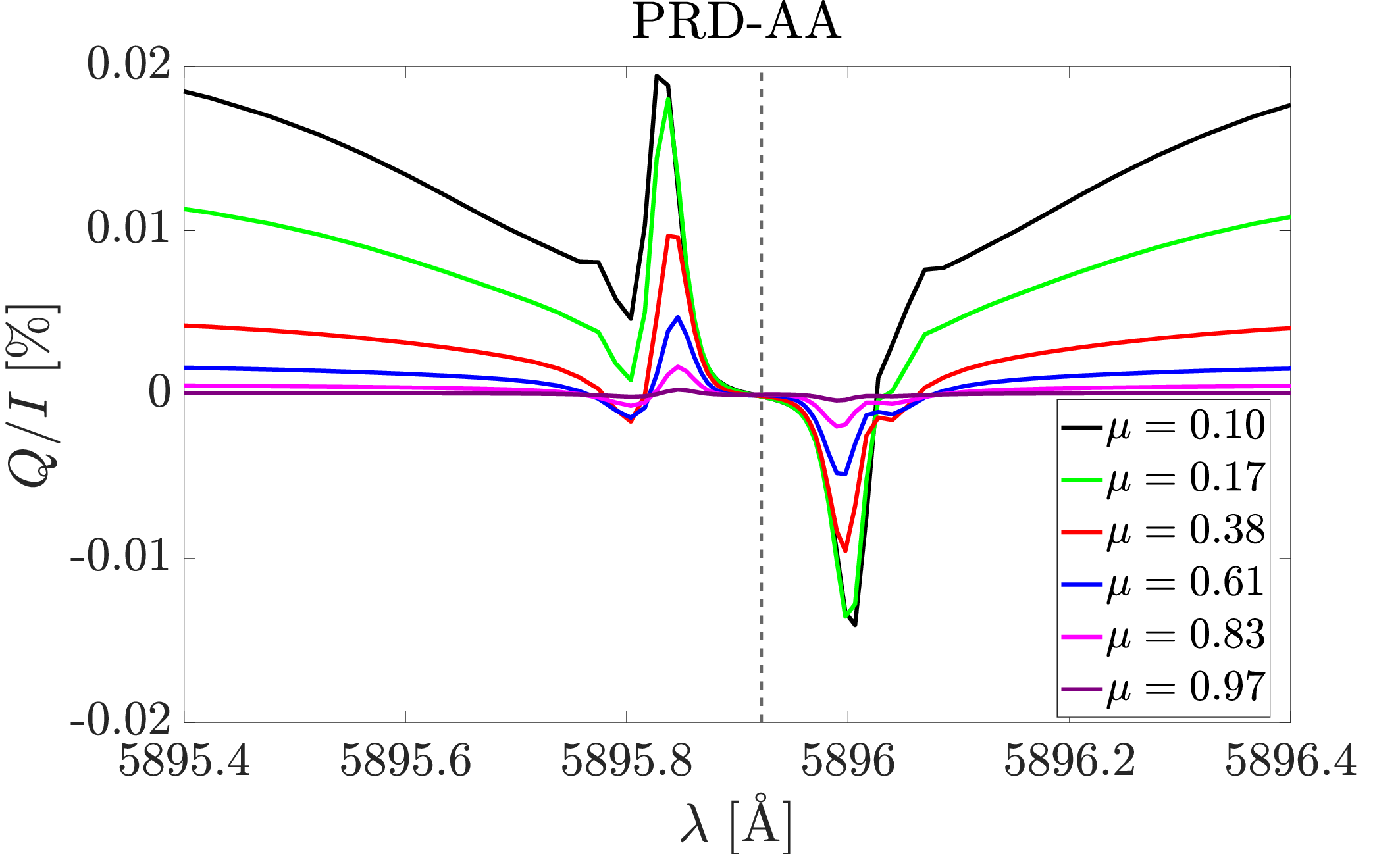}
    \caption{
    Emergent $Q/I$ profiles for the Na~{\sc{i}} D$_1$ line
    for different lines of sight, calculated 
    in atmospheric model C of \cite{fontenla1993},
    without a magnetic field.
    The results of calculations are shown taking PRD effects into account both in the general AD case (upper panel)
    and under the AA approximation (lower panel),
    in both cases neglecting the impact 
    of $J$-state interference.
    The vertical dashed line indicates the line-center wavelength.}
    \label{fig:PRD_AD_clv}
\end{figure}

Figure~\ref{fig:PRD_AD_clv} shows the same linear polarization profiles as in 
Figure~\ref{fig:B0}, also comparing AD and AA calculations, but for 
various lines of sight (LOS).
In the AA case (lower panel), the amplitude of 
the $Q/I$ peaks decreases for larger $\mu$, while its shape is largely preserved. 
In contrast, the shape of the AD profile (upper panel) clearly changes with
the LOS; the overall $Q/I$ pattern becomes much more symmetric for larger 
$\mu$ values, as the shift of the central peak towards the blue becomes 
less pronounced. In addition, the amplitude of the three peaks does not
decrease monotonically as $\mu$ increases, but instead they reach a maximum
and then decrease. The central peak, for instance, reaches its maximum at 
around $\mu = 0.61$.

\subsection{Impact of the magnetic fields on D$_1$ polarization}

\begin{figure}[ht!]
    \centering
    \includegraphics[width=0.49\textwidth]{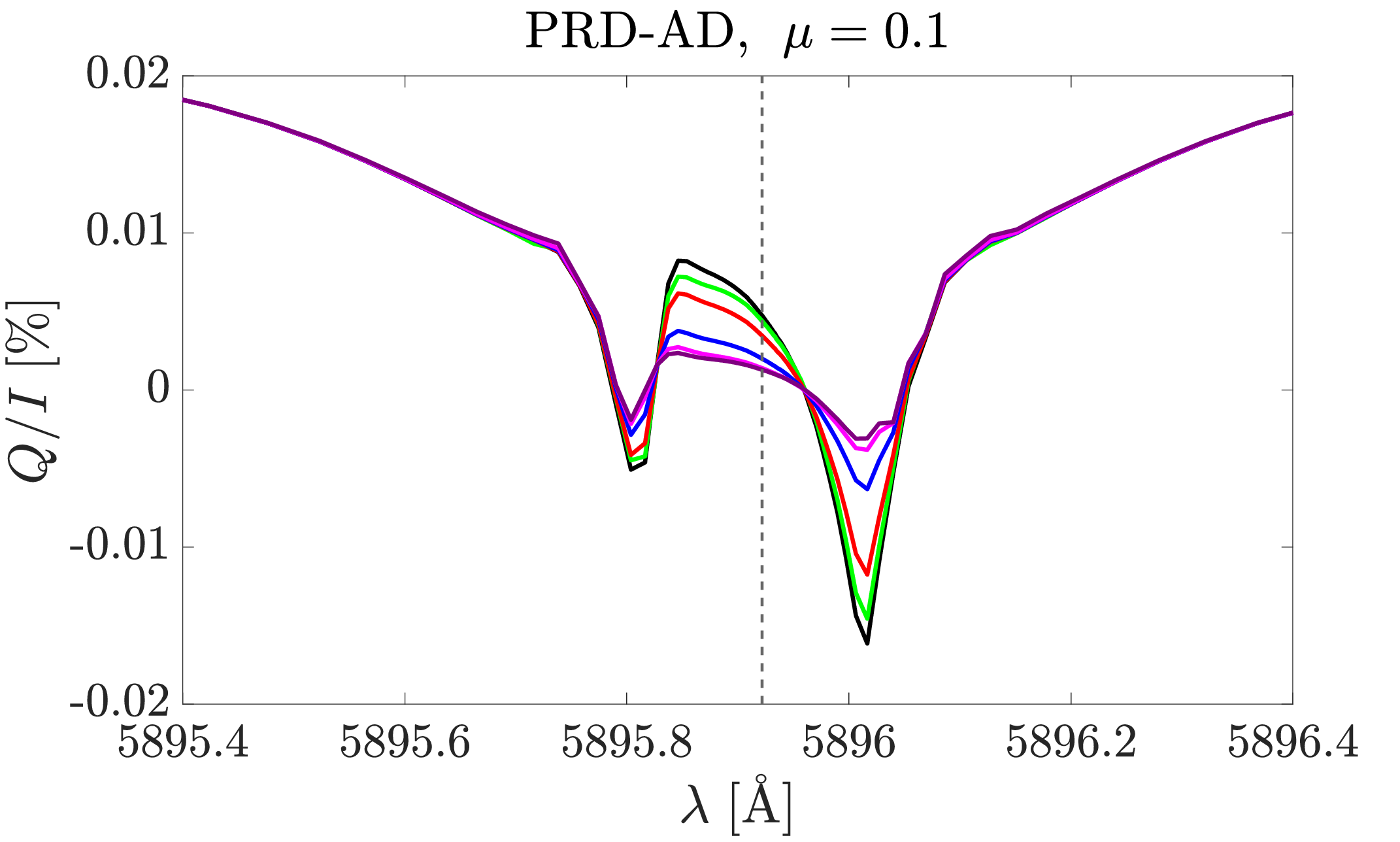}
    \includegraphics[width=0.49\textwidth]{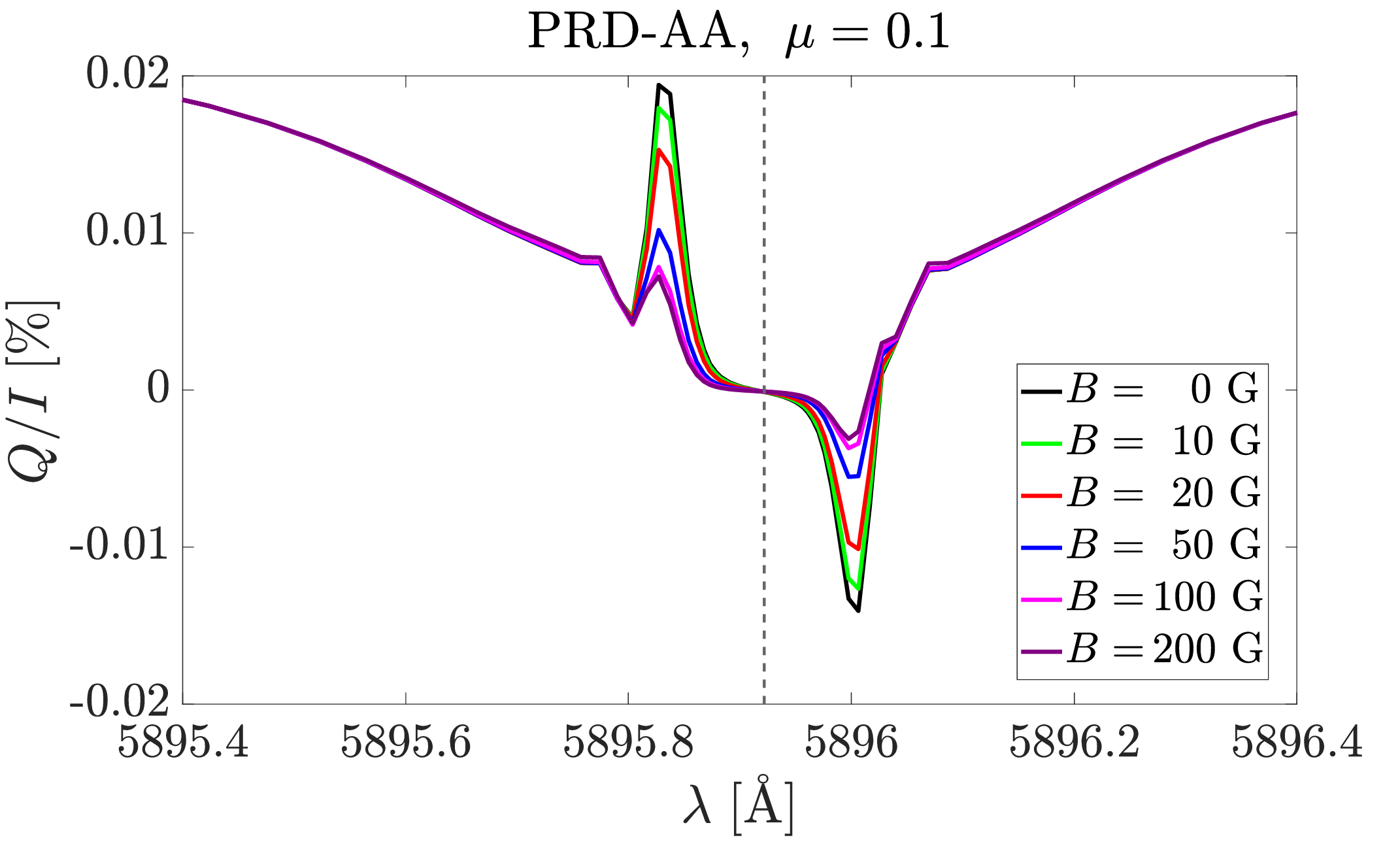}
    \caption{
    Emergent $Q/I$ profiles for the Na~{\sc{i}} D$_1$ line at $\mu = 0.1$, calculated 
    in atmospheric model C of \cite{fontenla1993},
    in the presence of a microturbulent magnetic field
    of various strengths,
    taking PRD effects into account both in the general AD case (upper panel)
    and under the AA approximation (lower panel),
    in both cases neglecting the impact 
    of $J$-state interference.
    The vertical dashed line indicates the line-center wavelength.}
    \label{fig:PRD_AD_B}
\end{figure}

It is also of interest to investigate the sensitivity of the D$_1$ scattering polarization to magnetic fields.
We recall that, in the presence of
an isotropic distribution of {micro-structured magnetic fields}, 
such as the one considered in this work, 
the Zeeman and magneto-optical signatures vanish due to cancellation effects, and the only impact of the magnetic field is the modification of $Q/I$ due to the Hanle effect \citep[see][]{alsina_ballester2021}.
The $U/I$ and $V/I$ signals
are thus zero and are consequently 
not shown.
The upper panel of Figure~\ref{fig:PRD_AD_B}
shows the $Q/I$ patterns obtained from AD calculations in which we accounted
for isotropic microturbulent magnetic fields of various strengths, 
in a range between $0$ 
and $100$~G, for which the 
incomplete Paschen-Back effect regime is attained.
For field strengths of $20$~G, the amplitude of the D$_1$ $Q/I$
signal already decreases appreciably due to the Hanle effect, both in the positive peak near the line 
center and in the negative lateral peaks.
At roughly $200$~G, Hanle saturation is reached and further increases in the field strength have no 
impact on the scattering polarization amplitude. 

Despite the differences in the shape of the $Q/I$ profile with respect to the PRD–AD case, the lower panel of Figure~\ref{fig:PRD_AD_B}
shows that the $Q/I$ peaks of the anti-symmetric PRD--AA profiles are sensitive to the same range of field strengths. 
Likewise, the decrease of the peak amplitudes at $100$~G, relative to the unmagnetized case, is similar to the one found for the AD case.   

\section{Conclusions}\label{sec:conclusions}

In the last years, theoretical investigations have demonstrated that the enigmatic 
linear polarization patterns observed in the core of the Na~{\sc{i}} D$_1$ line are a consequence of
the variation of the anisotropy of the radiation field in the wavelength interval spanned by
the line's various HFS components \citep[see][]{belluzzi2013,belluzzi2015,alsina_ballester2021}. 
Because of the high computational cost of the general PRD–AD calculations, the simplifying AA assumption was made in the works mentioned above.
Although this approximation is justified for modeling the intensity of spectral lines, it has been found to introduce inaccuracies in the core region
of the linear polarization profiles of certain strong resonance lines \citep[e.g.,][]{janett2021a}. 
In this work, we investigated the suitability of PRD--AA calculations for modeling the scattering 
polarization of the Na~{\sc{i}} D lines, using a numerical NLTE RT code for polarized 
radiation. The code is suitable for 1D atmospheric models and treats the spectral lines 
as arising from a
two-level atomic model with HFS (i.e., neglecting the impact of $J$-state interference), accounting for isotropic microturbulent magnetic fields
in the incomplete Paschen-Back effect regime. 

Considering the semi-empirical atmospheric model C of \cite{fontenla1993},
we first compared the D$_1$ linear polarization profiles obtained
under PRD--AA and PRD--AD calculations. In the absence of magnetic fields, the AD calculations
yield D$_1$ scattering polarization amplitudes of the order of $0.01\%$, 
while confirming that this signal originates from the detailed spectral structure
of the radiation field \citep[see][]{belluzzi2013}. 
However, the shape of the linear polarization profiles obtained under the PRD--AA and PRD--AD 
calculations is markedly different.
A largely antisymmetric two-peak profile is found in the AA 
case, with a positive peak to the blue of the D$_1$ line center and a negative peak to the red of it, 
whereas a far more symmetric three-peak profile is obtained from the PRD--AD calculation, 
with a positive peak 
near the line center and a negative peak at either side of it. 
The PRD--AD profile also exhibits an interesting behavior with the 
LOS. Its central peak, shifted towards the blue for $\mu = 0.1$, 
approaches the line center for larger $\mu$ values. As $\mu$ increases, 
the amplitudes of its three peaks grow until reaching a maximum and
then fall.

In the presence of isotropic microturbulent magnetic fields,
the sensitivity of the linear polarization profiles is quite similar
for the PRD--AA and PRD--AD calculations, at least in relative terms. 
The signal is appreciably depolarized in the presence of magnetic fields
of around $10$~G and, for stronger fields, its amplitude further reduces until about $200$~G, when saturation is reached. 
Nevertheless, it is worth noting that the various peaks obtained in the PRD--AD calculation
are depolarized to a slightly different extent by the magnetic field.

Some impact of PRD--AD effects is also appreciable in the D$_2$ line-core
$Q/I$ signal, where the dip feature found in the AA case is no longer obtained when such approximation is relaxed (see Figure~\ref{fig:PRD_AD_B_D2}).
Nevertheless, the relative impact on the linear polarization amplitude is quite minor 
and we find a very similar magnetic sensitivity for the PRD--AA and PRD--AD 
cases. Because the HFS separation of the upper level of the D$_2$ line is smaller than 
that of the upper level of D$_1$, the D$_2$ linear polarization is sensitive to weaker magnetic
fields; a depolarization is appreciable for strengths as low as $2$~G and saturation is found
beyond roughly $100$~G.

The remarkable influence of the general PRD--AD treatment
on the shape of the Na~{\sc{i}} D$_1$ scattering polarization pattern 
suggests that fully accounting for the angle-frequency
coupling is crucial in explaining 
spectro-polarimetric observations of this line, especially those that present
a rather symmetric $Q/I$ pattern.
Indeed, it should be expected that the astounding variety of 
linear polarization patterns that are observed in this line may be partly explained by 
phenomena that are intrinsically linked to the three-dimensional complexity of the solar atmosphere, 
including horizontal inhomogeneities in its thermodynamical properties, such as temperature or density,
and velocity gradients \citep[e.g.,][]{jaume_bestard2021}.
A very important goal 
for the near future is to model the D$_1$ and D$_2$
lines together, also accounting for PRD--AD effects and magnetic fields of arbitrary strength and orientation,
considering a two-term atomic model with HFS, thus considering also $J$-state interference.
This would allow us to perform a quantitative comparison with 
recent spectro-polarimetric observations, which revealed a wide variety of both nearly symmetric and antisymmetric $Q/I$ profiles of D$_1$ \citep[e.g.,][]{bianda2019}.
In any case, the results presented in this work bring us one step forward towards the goal of fully 
exploiting the Na~{\sc{i}} lines for diagnostics
of the elusive magnetic fields of the lower solar  chromosphere. 

\vspace{0.4cm}
\noindent \textit{Acknowledgements:} 
We gratefully acknowledge the financial support from the Sinergia program of the SNSF (grant No. CRSII5\_180238)
and from the European
Research Council (ERC) under the European Union’s Horizon 2020 research and innovation program (ERC Advanced
grant agreement No. 742265). T.P.A.'s participation in the publication is part of the Project RYC2021-034006-I,
funded by MICIN/AEI/10.13039/501100011033, and the European Union “NextGenerationEU”/RTRP.


\newpage

\appendix

\section{Redistribution matrix for a two-level atom with HFS}
\label{app:twolevel_HFS}

The redistribution matrix for a two-level atom with HFS 
in the incomplete Paschen-Back effect regime for the HFS is formally equivalent
to the one for a two-term atom (without HFS) in the Paschen-Back effect regime for FS, after making the appropriate
substitutions discussed below. 
However, after these substitutions, there are certain differences
between the atomic Hamiltonians of the two modelings 
in the presence of a magnetic field,
which impact the resulting eigenvalues and eigenvectors 
(i.e., the energies of the atomic states and their expansion coefficients, respectively). 
These differences are essential to accurately account for the magnetic sensitivity of 
the D$_1$ and D$_2$ lines. 

\subsection{Two-term atom}

When the magnetic splitting is comparable to the splitting between FS levels, the incomplete Paschen-Back 
effect regime
is attained and the total electronic
angular momentum $J$ is no longer a good quantum 
number for the atomic system.
The redistribution matrix (including both $R^{\scriptscriptstyle \rm II}$ 
and $R^{\scriptscriptstyle \rm III}$)
for a two-term atom in the Paschen-Back effect 
regime for the FS is given by Equation~(A1) of \citet{bommier2017}. 
Following the notation of that paper, the eigenvectors of the total 
Hamiltonian, which have the form $\ket{\beta L S J^\ast M}$,
can be expressed in
the basis of the eigenvectors of the total angular momentum as 
\citep[see Eq.~(3.58) of][]{landi_deglinnocenti+landolfi2004}
\begin{equation}
	\ket{\beta L S J^\ast M} = \sum_{J} C^J_{J^\ast M}(B) 
	\ket{\beta L S J M} , 
\end{equation}
where $\beta$ is a set of inner quantum numbers of the system, $L$ is the quantum number for orbital angular momentum, 
$S$ is the spin quantum number, 
and $M$ is the magnetic quantum number 
(i.e., the projection of $\bm{J}$ onto the selected quantization axis).
The modified number $J^\ast$ indicates the angular momentum number $J$
of the corresponding atomic state in the absence of magnetic fields. 
The energies of the levels $\ket{\beta L S J^\ast M}$ and the expansion 
coefficients $C^J_{J^\ast M}(B)$ are calculated through the diagonalization
of the atomic Hamiltonian. Its matrix elements are 
\begin{equation}
	\bra{\beta L S J M}H_{\rm{so}} + H_B\ket{\beta L S J' M},
\end{equation}
where $H_{\rm{so}}$ and $H_B$ are the spin-orbit and magnetic Hamiltonian, 
respectively.
For the magnetic fields typically found in astrophysical plasmas
one can take $H_B = \mu_0 \, (\bm{J} + 2\bm{L}) \, \bm{B}$,
where $\mu_0$ is the Bohr magneton.
By assuming that the atomic system is described by L--S coupling and 
applying the Wigner-Eckart theorem and its corollaries,
one finds that the only 
non-zero matrix elements are
\citep[see 
Equations~(3.61a) and (3.61b) of][]{landi_deglinnocenti+landolfi2004}
\begin{equation}
	\bra{\beta L S J M}H_{\rm{so}} + H_B\ket{\beta L S J M} = 
	E_{\beta L S}(J) + \mu_0 B g_{LS}(J) M , 
  \label{eq:ap_JelementDiag} 
\end{equation}
where $E_{\beta L S}(J)$ is the energy of the $J$-level and $g_{LS}(J)$ its 
Land\'e factor
\citep[see 
Equation~(3.8) of][]{landi_deglinnocenti+landolfi2004}
\begin{equation}\label{eq:gLS}
	g_{LS}(J) = 1 + \frac{1}{2} \frac{J(J+1) +S(S+1) -L(L+1)}{J(J+1)} 
	\qquad (J \ne 0) ,
\end{equation}
and
\begin{align}
	& \bra{\beta L S J-1 \, M}H_{\rm{so}} + H_B\ket{\beta L S J M} = 
	\bra{\beta L S J M} H_{\rm{so}} + H_B \ket{\beta L S J-1 \, M} 
	\nonumber \\ 
	& = - \frac{\mu_0 B}{2J} \sqrt{\frac{(J+S+L+1)(J-S+L)(J+S-L)(-J+S+L+1)
	(J^2-M^2)}{(2J+1)(2J-1)}}.
 \label{eq:ap_JelementTriang}
\end{align} 

\subsection{Two-level atom with HFS}

The redistribution matrix for a two-level atom with HFS is still given by 
Equation~(A.1) of \citet{bommier2017},
with the following formal substitutions
\citep[see Equation~(7.64) of][]{landi_deglinnocenti+landolfi2004}
\begin{align}
	& \beta \rightarrow \alpha \qquad S \rightarrow I \qquad  
	J^\ast \rightarrow F^\ast \nonumber \\
	& L \rightarrow J \qquad  J \rightarrow F \qquad M \rightarrow f,
 \label{eq:FS_to_HFS}
\end{align}
where $\alpha$ is a set of inner quantum numbers that includes $S$ and $L$, $I$ is the quantum number for nuclear spin, $F$ is the quantum number for total atomic angular momentum such that $\bm{F} = \bm{J} + \bm{I}$, 
and $f$ is the projection of $\bm{F}$ upon the selected quantization axis. 
The number $F^\ast$ indicates the angular momentum number $F$ of the 
{corresponding} atomic state in the absence of magnetic fields.
The eigenvectors of the total Hamiltonian, which have the form 
$\ket{\alpha J I F^\ast f}$, can be expanded on the basis of the eigenvectors 
of the total angular momentum as
\citep[see Equation~(7.58) of][]{landi_deglinnocenti+landolfi2004}
\begin{equation}
	\ket{\alpha J I F^\ast f} = \sum_{F} C^F_{F^\ast f}(B) 
	\ket{\alpha J I F f}.
\end{equation}
In this work, we calculate the energies of the levels $\ket{\alpha J I F^\ast f}$ and the expansion coefficients $C^F_{F^\ast f}(B)$ 
through the diagonalization of the atomic Hamiltonian, whose elements are
\begin{equation}
	\bra{\alpha J I F f} H_{\rm{HFS}} + H_B\ket{\alpha J I F' f},
\end{equation}
where $H_{\rm{HFS}}$ and $H_B$ are the HFS and magnetic Hamiltonian, 
respectively. 
The contribution from $H_{\rm{HFS}}$ is obtained following 
Equation~(3.70) of \citet{landi_deglinnocenti+landolfi2004}.
As in the case for the two-term atom, 
the contribution from $H_B$ is obtained from the application of 
the Wigner-Eckart theorem and its corollaries
\citep[see Equations~(3.71) and (3.73) of][]{landi_deglinnocenti+landolfi2004}.
One finds that the only non-zero matrix elements are 
\begin{equation}
	\bra{\alpha J I F f}H_{\rm{HFS}} + H_B\ket{\alpha J I F f} = 
	E_{\alpha J I}(F) + \mu_0 B \, g_{LS}(J) \, g_{\rm{HFS}}(F) \, f , 
  \label{eq:ap_FelementDiag}
\end{equation}
where
\begin{equation}
	E_{\alpha J I}(F) = E_{\beta L S}(J) + 
	\frac{\mathcal{A}(\alpha, J, I)}{2} K + 
	\mathcal{B}(\alpha, J, I) \left[ K(K+1) - \frac{4}{3}J(J+1)I(I+1)
	\right],
\end{equation}
with $\mathcal{A}$ and $\mathcal{B}$ the HFS constants, 
$K=F(F+1)-J(J+1)-I(I+1)$, 
\begin{equation}\label{eq:gHFS}
	g_{\rm{HFS}}(F) = \frac{1}{2} \frac{F(F+1) + J(J+1) - I(I+1)}{F(F+1)}
	\qquad (F\ne0),
\end{equation}
and
\begin{align}
	\bra{\alpha J I F-1 \, f}&H_{\rm{HFS}} + H_B\ket{\alpha J I F f} = 
	\bra{\alpha J I F f}H_{\rm{HFS}} + H_B\ket{\alpha J I F-1 \, f} 
	\nonumber \\
	& = \mu_0 B g_{LS}(J) (-1)^{J+I-f} \sqrt{J(J+1)(2J+1)(2F+1)(2F-1)} 
	\left\{
	\begin{array}{ccc}
		F-1 & F & 1 \\
		J & J & I 
	\end{array}
	\right\}
	\left(
	\begin{array}{ccc}
		F & F-1 & 1 \\
		-f & f & 0 
	\end{array}
	\right) \nonumber \\
	& = \frac{\mu_0 B g_{LS}(J)}{2F} 
	\sqrt{\frac{(F+J+I+1)(F-I+J)(F+I-J)(-F+J+I+1)(F^2-f^2)}{(2F+1)(2F-1)}}.
 \label{eq:ap_FelementTriang}
\end{align}
%
In this work, we compute the eigenvectors and eigenvalues of the atomic 
Hamiltonian by considering the matrix elements given in 
Equations~\eqref{eq:ap_FelementDiag} and \eqref{eq:ap_FelementTriang}. 
We note that these elements are not equivalent to the analogous ones 
for the two-term atom without HFS obtained via the formal 
substitutions shown in Equations~\eqref{eq:FS_to_HFS}.
The contribution from the magnetic Hamiltonian in 
Equation~\eqref{eq:ap_FelementDiag} differs from the one that would be 
obtained from Equation~\eqref{eq:ap_JelementDiag} in the 
$g_{LS}(J) \, g_{\rm{HFS}}(F)$ factor of the former. In addition, the element 
in Equation~\eqref{eq:ap_FelementTriang} differs from the one that would be 
obtained from Equation~\eqref{eq:ap_JelementTriang} in the overall sign and 
in the $g_{LS}(J)$ factor.

\section{Impact of PRD--AD treatment on D$_2$ polarization}
\label{app:D2}

\begin{figure}[ht!]
    \centering
    \includegraphics[width=0.49\textwidth]{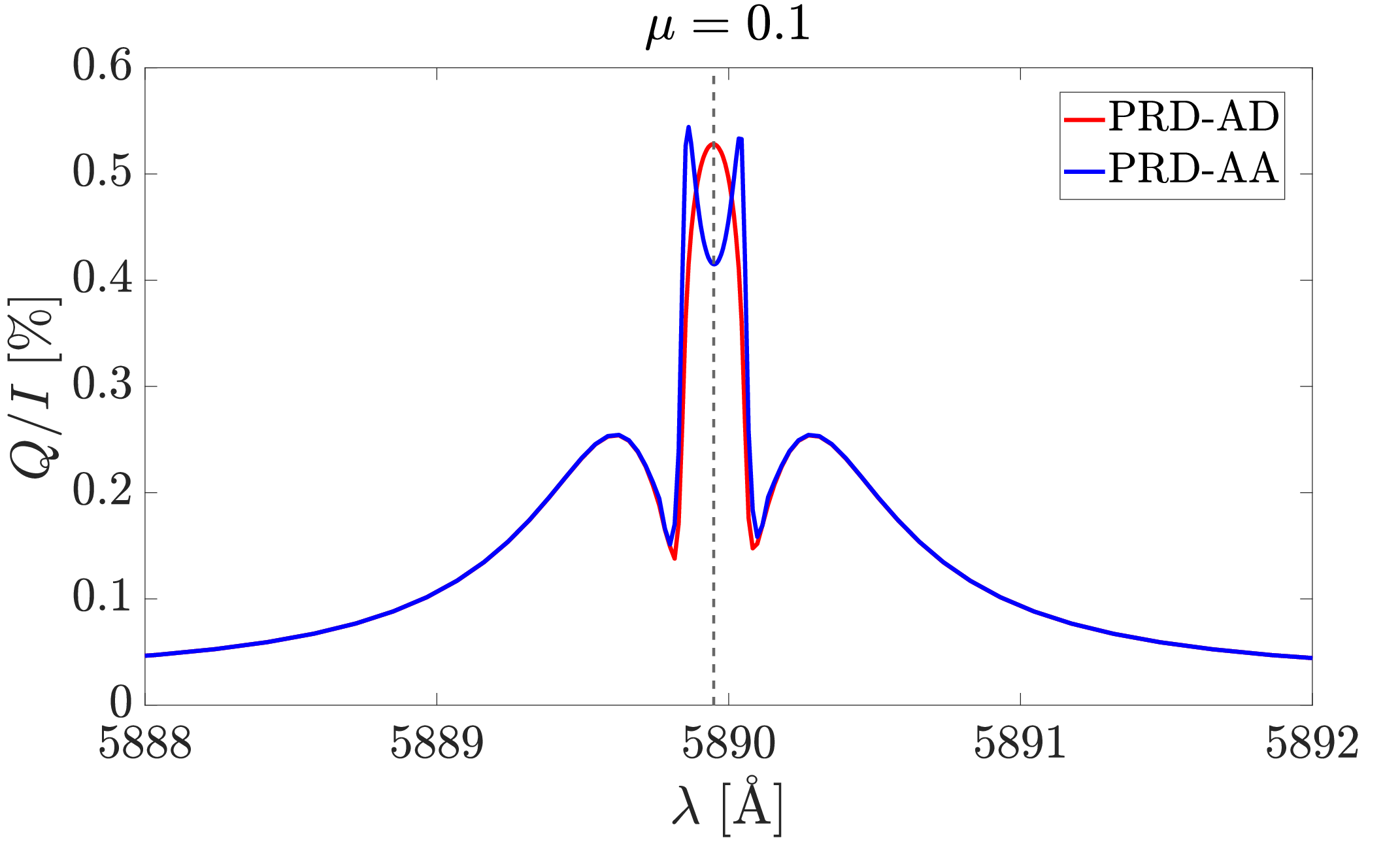}
    \caption{
    Emergent $Q/I$ profiles for the Na~{\sc{i}} D$_2$ line at $\mu = 0.1$, calculated with a two-level model atom and in the atmospheric model C of \cite{fontenla1993}, without a magnetic field.
    The results of calculations are shown
    taking PRD effects into account both in the general AD case (red curve)
    and under the AA approximation (blue curve).
    The vertical dashed line indicates the line-center wavelength.
    }
    \label{fig:B0_D2}
\end{figure}

\begin{figure}[ht!]
    \centering
    \includegraphics[width=0.48\textwidth]{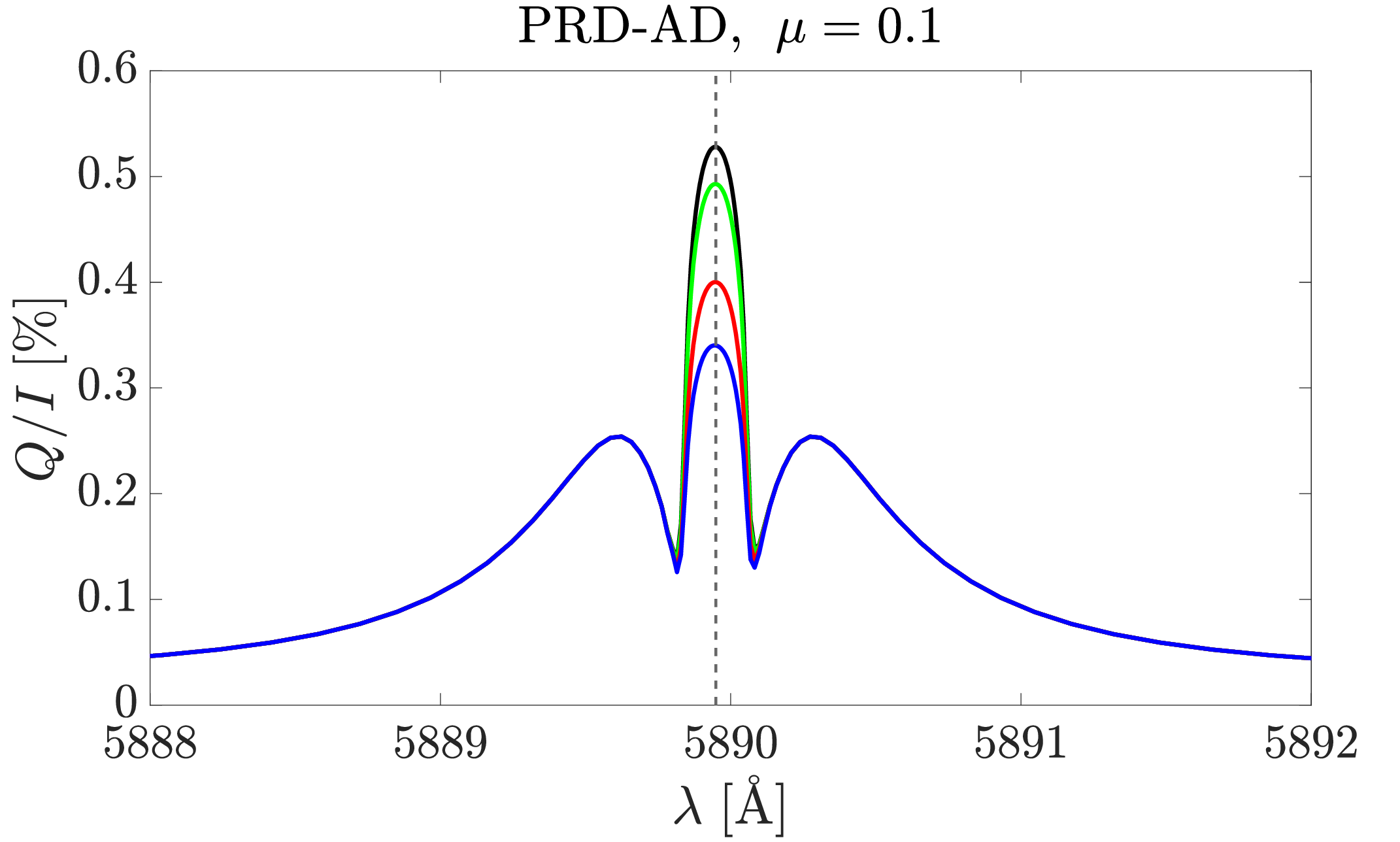}
    \includegraphics[width=0.48\textwidth]{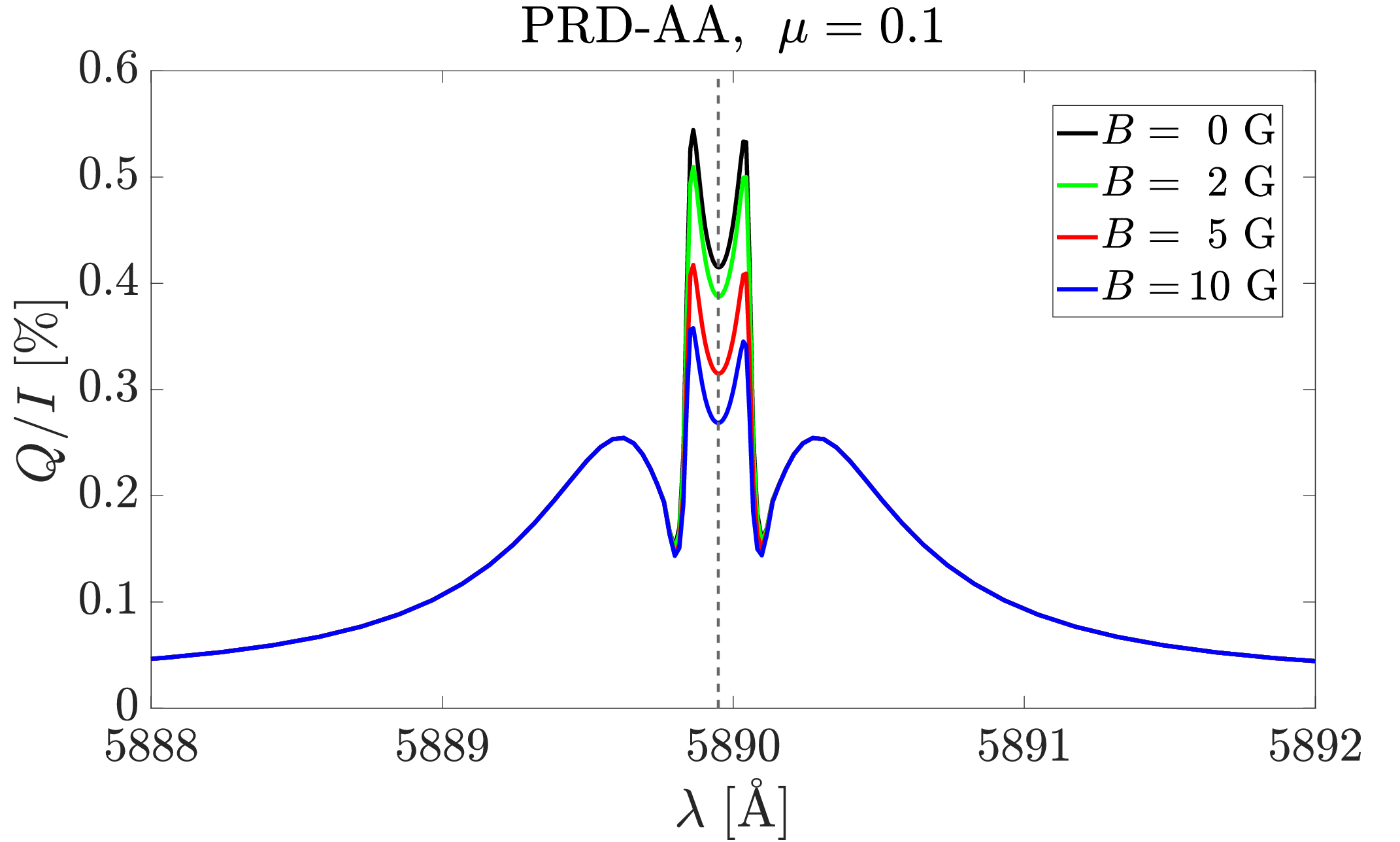}
    \caption{
    Emergent $Q/I$ profiles for the Na~{\sc{i}} D$_2$ line at $\mu = 0.1$, calculated 
    in atmospheric model C of \cite{fontenla1993},
    in the presence of a microturbulent magnetic field
    of various strengths,
    taking PRD effects into account both in the general AD case (left panel)
    and under the AA approximation (right panel). 
    The vertical dashed line indicates the line-center wavelength.
    }
    \label{fig:PRD_AD_B_D2}
\end{figure}

In this section, we analyze the suitability of a PRD--AA modeling for the Na~{\sc{i}} D$_2$ line at 5890~{\AA}, through quantitative comparisons with the general PRD--AD calculations, both in the absence and presence of magnetic fields.
We considered the same setting exposed in Section~\ref{sec:numerical_setting}
and the wavelength interval
$[\lambda_{\min},\lambda_{\max}]=[5887.3\text{\,\AA},5892.6\text{\,\AA}]$.
Figure~\ref{fig:B0_D2} compares
the $Q/I$ patterns of the D$_2$ line obtained for PRD--AA and 
for PRD--AD calculations at $\mu = \cos(\theta) = 0.1$ and in the absence of magnetic fields.
The left and right panels of Figure~\ref{fig:PRD_AD_B_D2} show the $Q/I$ patterns obtained from AD and AA calculations, respectively, at $\mu = \cos(\theta) = 0.1$.
In these calculations, we accounted for isotropic microturbulent magnetic fields of various strengths, ranging from $0$ to $10$~G, 
for which the incomplete Paschen-Back effect regime must be considered. 
Under the PRD--AA approximation, the $Q/I$ profile of the D$_2$ line 
presents an artificial dip in its core,
which is not found for the PRD--AD case.
This result is similar to what was reported in an analogous investigation on the Ca~{\sc{i}} line at $4227$~\AA\ by \citet{janett2021a}.
However, the relative impact on the linear polarization amplitude is quite minor when compared to the Na~{\sc{i}} D$_1$ case.
Moreover, we find a very similar magnetic sensitivity for the PRD--AA and PRD--AD cases.

\section{Angular sampling}\label{app:angular_grid}

\begin{figure}[ht!]
    \centering
    \includegraphics[width=0.48\textwidth]{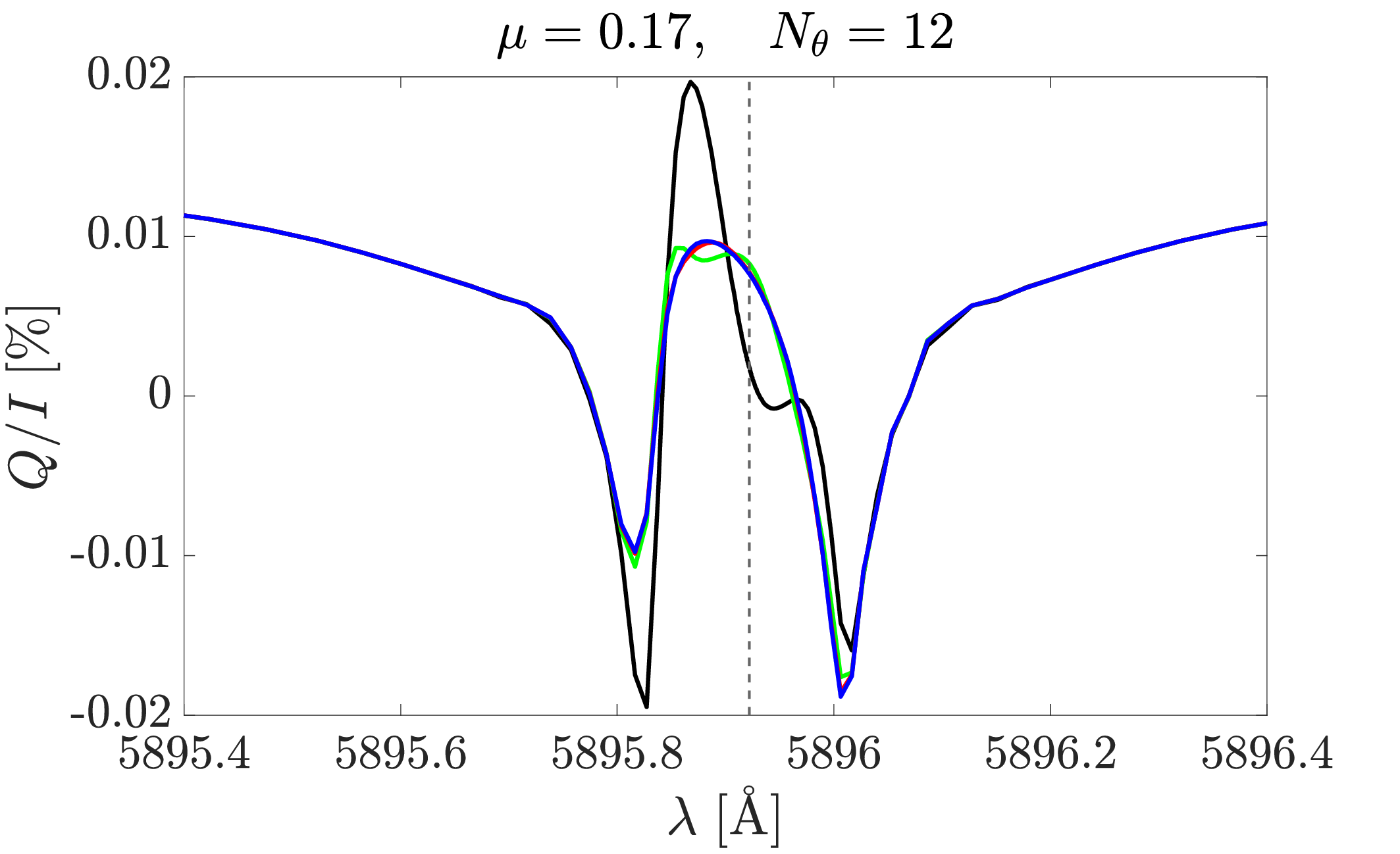}
    \includegraphics[width=0.48\textwidth]{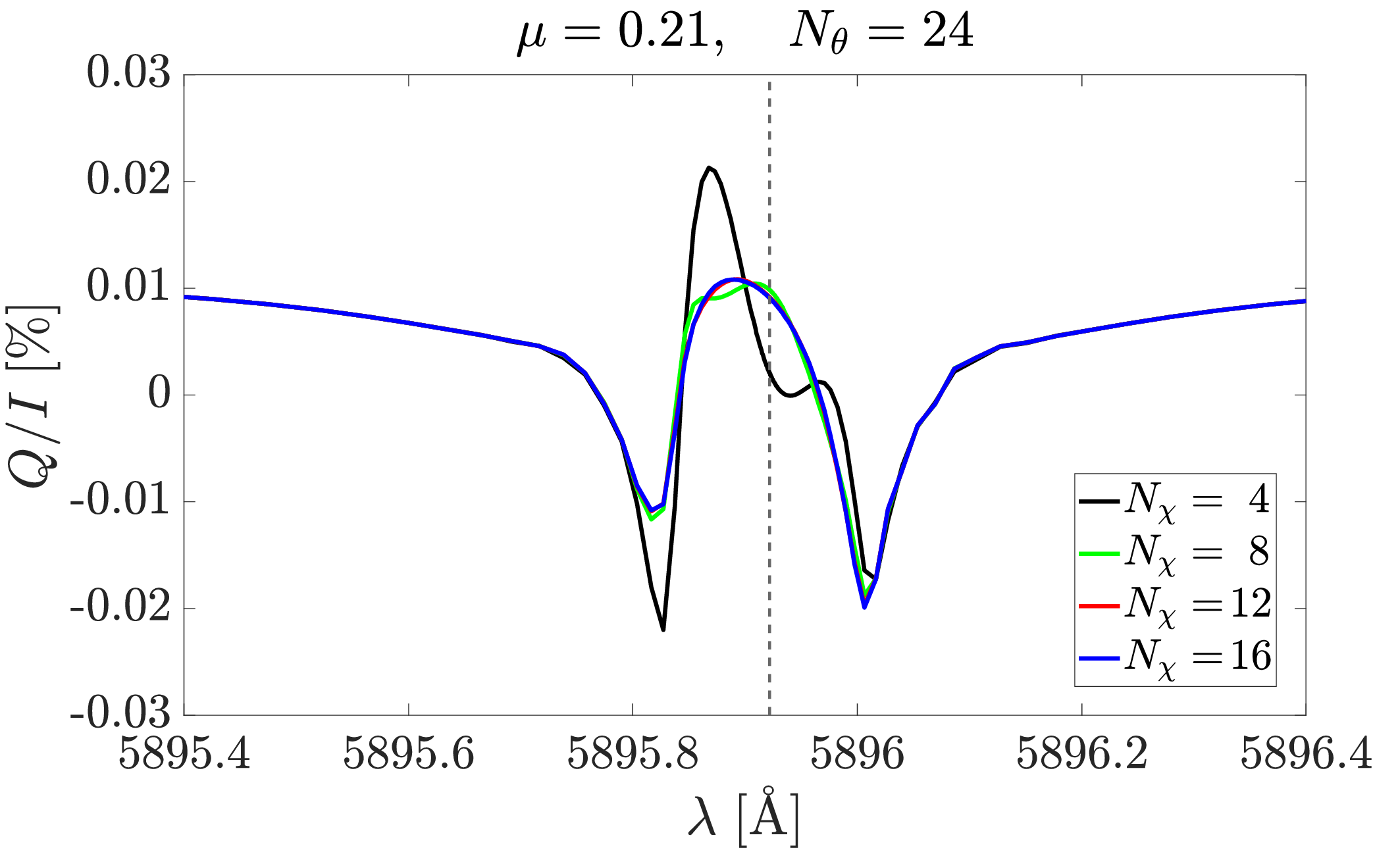}
    \vspace*{-0.1cm}
    \caption{Emergent $Q/I$ profiles for the Na~{\sc{i}} D$_1$ line
    obtained through the PRD calculations detailed in the main 
    text in the general AD case. Model C of \cite{fontenla1993}
    was considered in the absence of magnetic fields.
    The vertical dashed line indicates the line-center wavelength.
    The results of calculations are shown for angular grids 
    with $N_\chi=4,8,12,16$ (see legend) and
    with $N_\theta = 12$ at $\mu = 0.17$ (left panel)
    and $N_\theta = 24$ at $\mu = 0.21$ (right panel).  
    All the results shown in this work were thus obtained considering $N_\chi=12$ and $N_\theta = 12$.}
    \label{fig:azimuth_sampling}
\end{figure}

For the angular discretization of $\bm{\Omega}=(\theta,\chi)$, we use a tensor product quadrature.
For the inclination $\mu=\cos(\theta)\in[-1,1]$, we consider two Gauss-Legendre grids (and corresponding weights)
for $\mu\in(-1,0)$ and $\mu\in(0,1)$, respectively, with  $N_\theta/2$ nodes each, namely
$$ 1 > \mu_1 > \mu_2 > \cdots > \mu_{N_{\theta}/2} > 0 >  \mu_{N_{\theta}/2 + 1} > \cdots > \mu_{N_{\theta}} > -1.$$
This grid corresponds to
$$ 0 < \theta_1 < \theta_2 < \cdots < \theta_{N_{\theta}/2} < \pi/2 <  \theta_{N_{\theta}/2 + 1} < \cdots < \theta_{N_{\theta}} < \pi,$$
with $\theta_j=\arccos(\mu_j)$ for $j=1,\ldots,N_\theta$, with $N_\theta$ even. For the azimuth $\chi\in(0,2\pi]$,
we consider an equidistant grid (and corresponding trapezoidal weights) with $N_\chi$ nodes, namely
$$\chi_k = k\cdot 2\pi/N_\chi \quad \text{for} \quad k=1,\ldots,N_\chi.$$
%
Figure~\ref{fig:azimuth_sampling} presents
the emergent $Q/I$ profiles for the Na~{\sc{i}} D$_1$ line
obtained with PRD--AD calculations 
in the absence of magnetic fields
with $N_\theta = 12$ at $\mu = 0.17$ (left panel)
and $N_\theta = 24$ at $\mu = 0.21$ (right panel)
for various azimuthal samplings. 
The two panels are shown for different LOSs, so that they 
lie on the considered angular grid.
For both $N_\theta = 12,24$,
the very common azimuthal sampling 
given by $N_\chi=8$
was revealed to be inaccurate for modeling these scattering polarization signals,
taking PRD effects into account in the general AD case.
Therefore, we set $N_\chi=12$ in all the AD calculations shown in this work.

\vspace*{-0.2cm}

\newpage

\section{Frequency-integrated D$_1$ linear polarization signal}\label{app:QI_integrated}

\begin{figure}[ht!]
    \centering
    \includegraphics[width=0.48\textwidth]{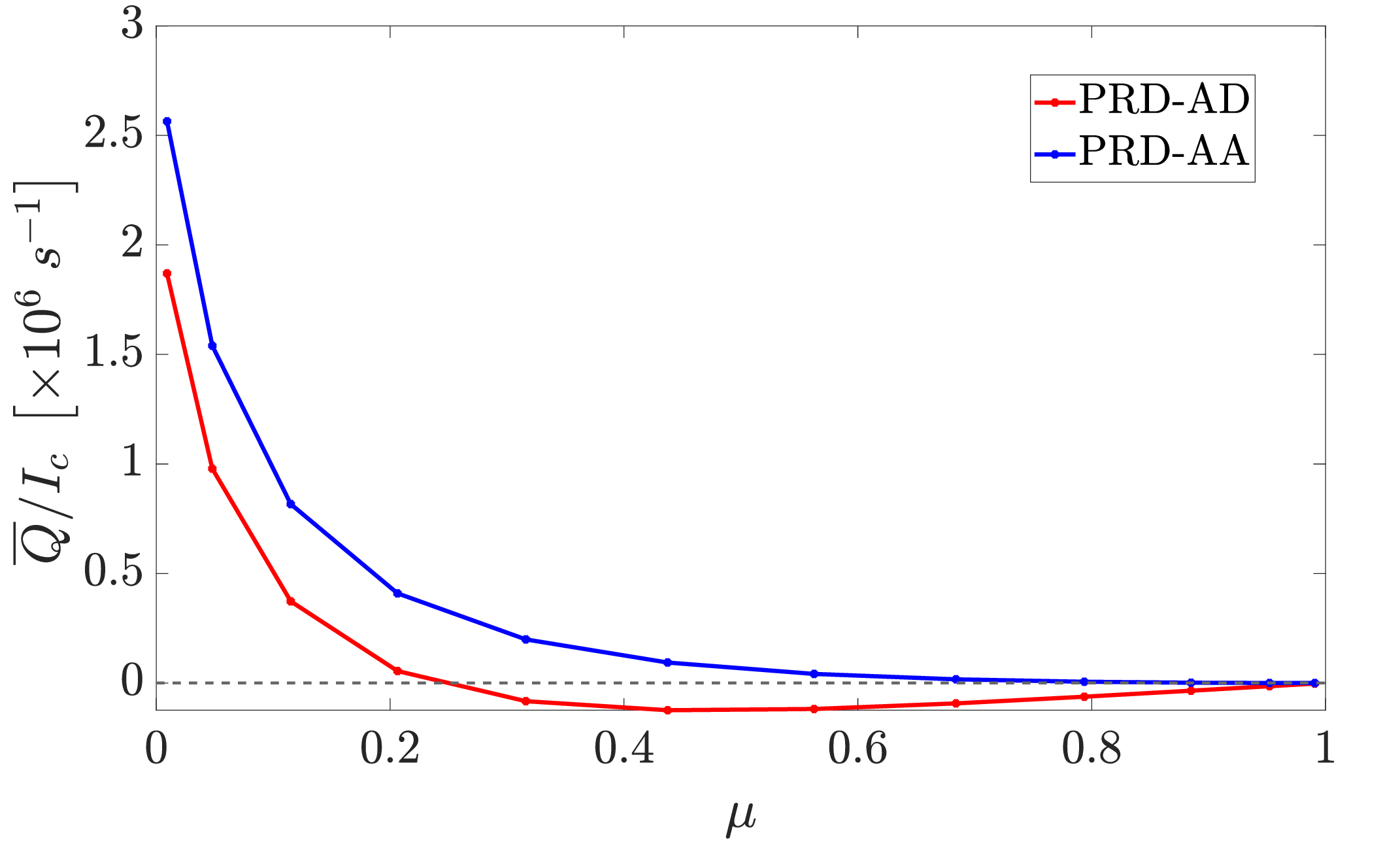}
    \caption{Frequency-integrated emergent $\overline Q/I_c$ signals as a function 
    of $\mu$ for the Na~{\sc{i}} D$_1$ line, calculated in atmospheric model C of \cite{fontenla1993} in the absence of magnetic fields. 
     The signals were obtained from the profiles calculated neglecting the continuum contribution to polarization and
     taking PRD effects into account 
     both in the general AD case (red curve) and under the AA approximation (blue curve), for an angular grid with $N_\chi=12$ and $N_\theta=24$.
    }
    \label{fig:QI_integrated}
\end{figure}

Figure~\ref{fig:QI_integrated} shows the frequency-integrated $\overline Q$ signals of the Na~{\sc{i}} D$_1$ line, normalized to the continuum
intensity $I_c$, as a function of $\mu$. It presents a comparison between the signals obtained from calculations taking PRD effects into account in
the general AD case and under the AA approximation
in the absence of magnetic fields, neglecting the continuum contribution to polarization.
We note that the amplitudes in the AD curve are comparable to those in the AA one,
indicating that the AD treatment does not lead to an enhancement of the net $\overline Q/I_c$ signal.
In addition, 
we computed the emergent $\overline Q/\overline I$ ratio 
in a frequency interval corresponding
to $\pm 0.50$~\AA\ from the line center.\footnote{We note that, neglecting continuum polarization, $\overline Q$
remains constant for a larger frequency integration interval, while $\overline I$ increases.}
We found the largest values for $\overline Q/\overline I$ to be on the order of $10^{-5}$, for $\mu$ positions close to the limb.
The largest contribution to the net linear polarization comes from the regions between $\sim\!0.15$ and $\sim\! 0.40$~\AA\ from the line center,
both to the red and the blue, where Stokes $Q$ signals of the same sign are found. Such wing signals arise mainly as a consequence of transfer 
effects. To verify this, we computed the
$\overline\varepsilon_Q/\overline\varepsilon_I$
ratio for the AA case, neglecting continuum contributions,
at all height points between $600$ and $2017$~km in the FAL-C atmospheric model. 
The largest value of $\overline\varepsilon_Q/\overline\varepsilon_I$,
found for $\mu$ positions close to the limb,
is on the order of $10^{-6}$, that is one order of magnitude smaller than the 
emergent $\overline Q/\overline I$ ratio.

\bibliographystyle{aa}
\bibliography{bibfile}

\end{document}